\documentclass[sigconf]{acmart}
\usepackage{enumitem}
\usepackage{multirow}

\AtBeginDocument{%
  }
    
\setcopyright{acmlicensed}
\copyrightyear{2018}
\acmYear{2018}
\acmDOI{XXXXXXX.XXXXXXX}
\acmConference[Conference acronym 'XX]{Make sure to enter the correct
  conference title from your rights confirmation email}{June 03--05,
  2018}{Woodstock, NY}
\acmISBN{978-1-4503-XXXX-X/2018/06}

\begin{document}

\title{A Joint Auction Framework with Externalities and Adaptation}


\author{Chun Fang$^{\dagger,*}$, Luowen Liu$^{*}$, Kun Huang$^{*}$, Tao Ruan, Sheng Yan, Zhen Wang \\
Huan Li, Qiang Liu, Xingxing Wang}
\affiliation{
  \institution{Meituan}
  \city{Beijing}
  \country{China}
}

\email{{fangchun02, liuluowen, huangkun13, ruantao02, yansheng05, wangzhen179}@meituan.com}
\email{{lihuan70, liuqiang43, wangxingxing04}@meituan.com}

\thanks{$\dagger$ Corresponding author.}
\thanks{*Both authors contributed equally to this research.}
\renewcommand{\shortauthors}{Chun Fang et al.}





\begin{abstract}
Recently, joint advertising has gained significant attention as an effective approach to enhancing the efficiency and revenue of advertising slot allocation. Unlike traditional advertising, which allocates advertising slots exclusively to a single advertiser, joint advertising displays advertisements from brands and stores that have established a joint selling relationship within the same advertising slot. However, existing approaches often struggle to accommodate both joint and traditional advertising frameworks, thereby limiting the revenue potential and generalizability of joint advertising. Furthermore, these methods are constrained by two critical limitations: they generally neglect the influence of global externalities, and they fail to address the bidding variability stemming from multi-party advertiser participation. Collectively, these limitations present substantial challenges to the design of joint auction mechanisms. To address these challenges, we propose a \textbf{J}oint Auction Framework incorporating \textbf{E}xternalities and \textbf{A}daptation, and leverage the automated mechanism design (AMD) method through our proposed \textbf{JEANet} to compute joint auction mechanisms that satisfy the conditions of individual rationality (IR) and approximate dominant strategy incentive compatibility (DSIC). As the first AMD method to integrate global externalities into joint auctions, JEANet dynamically adapts to the bidding characteristics of multi-party advertiser and enables unified auctions that integrate both joint and traditional advertising. Extensive experimental results demonstrate that JEANet outperforms state-of-the-art baselines in multi-slot joint auctions.

\end{abstract}





\maketitle


\section{Introduction}
Advertising is a key revenue source for Internet companies \cite{edelman2007internet,qin2015sponsored}, directly impacting company development and Internet prosperity. Optimizing ad revenue is thus a research focus in academia and industry. In online advertising, platforms usually determine ad slot allocations and payment rules, with payments based on advertiser bids and auction mechanisms.

In traditional advertising, each page browsing request from a user triggers the display of an ordered list of products \cite{yan2020ads,zhang2018whole}. When a user clicks on an ad, the platform charges the corresponding advertiser a fee, and advertisers can raise their bids to secure better display positions. Now, brand suppliers (not just stores) seek to participate in ad slot auctions to enhance awareness. Hence, "joint advertising" has emerged, enabling stores and brands to co-participate via joint selling—each ad slot featuring a store-brand bundle \cite{ma2024joint,zhang2024joint}. Compared with traditional advertising, it satisfies both parties’ needs, boosts platform revenue, and fosters win-win outcomes—now adopted by major e-commerce platforms \cite{aggarwal2024selling}.

Whether in traditional or joint advertising, platforms display ads alongside organic items from recommendation systems. Though organic items generate no direct revenue, they enhance user experience, boosting engagement and retention, which contribute to the platform's long-term value. Thus, hybrid lists must balance ads and organic items to reconcile revenue with user experience  \cite{ma2025context}. Designing ad mechanisms that account for externalities has been a key academic and industry focus. Traditional ad research has evolved from local to global perspectives. Local models have explored factors such as user clicks \cite{ghosh2008externalities}, ad candidate sets \cite{liu2021neural}, and ad positions \cite{huang2021deep}. Several studies \cite{liao2022cross,wang2019learning,xie2021hierarchical} investigated externality-aware ad slot allocation but overlooked ads-organic items interplay, limiting global hybrid ranking optimization. Recent works \cite{gatti2012truthful,liao2022nma} explored ads-organic items global interactions but are inapplicable to joint advertising, failing to account for store-brand joint selling relationship. As far as we know, no existing auction mechanism adequately addresses the externality problem in joint advertising.

\begin{figure}[H]
  \centering
  \includegraphics[width=\linewidth]{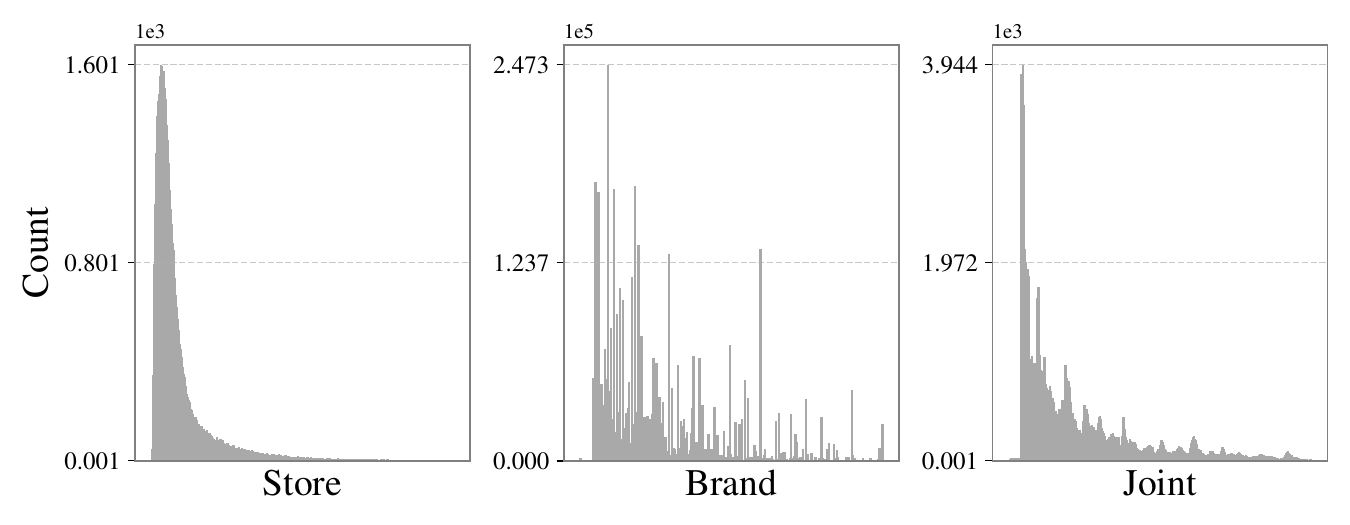}
  \caption{Histogram of bid differences among store, brand, and joint advertisers. For comparison, this figure illustrates the normalized bid distribution of different advertisers.}
  \label{fig:distribution}
  \Description{Histogram of bid differences among store, brand, and joint advertisers. For comparison, this figure illustrates the normalized bid distribution of different advertisers.}
\end{figure}

Compared with traditional advertising auctions, joint auctions are not merely a simple overlay of supply. Instead, they must account for the more complex four-party interaction among stores, brands, platforms, and users, as well as the joint selling relationships established by stores and brands based on their marketing objectives. As illustrated in \autoref{fig:distribution}, advertisers' bids often exhibit distinct distributions due to variations in their marketing objectives and the private values of advertisements. Moreover, in real-world business scenarios, joint advertising significantly reduces the size of the candidate set compared to traditional advertising under the same supply conditions. This reduction impacts advertising effectiveness and directly limits the practical application and promotion of joint advertising.

\begin{figure}
  \centering
  \includegraphics[width=\linewidth]{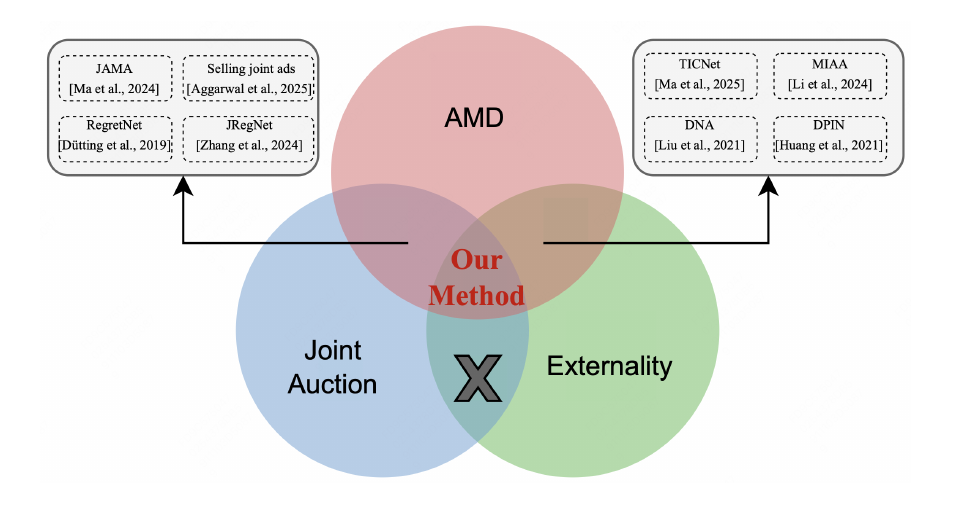}
  \caption{JEANet is the first AMD method that explicitly accounts for externalities in joint auctions. Numerous studies have employed the AMD method to analyze externalities in traditional and joint advertising. However, no research has explored the impact of externalities in joint advertising.}
  \label{fig:3-feature relationship}
  \Description{JEANet is the first AMD method that explicitly accounts for externalities in joint auctions. Numerous studies have employed the AMD method to analyze externalities in traditional and joint advertising. However, no research has explored the impact of externalities in joint advertising.}
\end{figure}

In recent years, with the advent of differentiable economics \cite{dutting2024optimal}, which leverages deep learning to discover optimal mechanisms, AMD has emerged as a new paradigm for multi-slot auction design \cite{conitzer2002complexity,conitzer2004self}. This approach offers a potential framework for addressing the challenges in joint advertising. As illustrated in \autoref{fig:3-feature relationship}, existing AMD methods that consider externalities are ill-suited for joint advertising, as they fail to model joint selling relationships. Conversely, AMD methods applicable to joint advertising neglect the effects of global externalities and bid heterogeneity, resulting in hybrid lists that lack global optimality. Therefore, researching a joint auction mechanism that accounts for externalities, adapts to bid distributions, and integrates seamlessly with traditional advertising holds substantial practical value.

\subsection{Main Contributions}
To address these challenges, we propose a joint auction framework integrating externalities and adaptation, aiming to optimize platform revenue and user experience while satisfying IR and approximate DSIC. Within this framework, ads and organic items are jointly allocated and priced to form a globally optimal hybrid list. We leverage the AMD methods to compute the joint auction mechanism by proposing the \textbf{J}oint Auction with \textbf{E}xternalities and \textbf{A}daptation \textbf{Net}work (\textbf{JEANet}). JEANet flexibly allocates slots to store, brand, or joint advertisers, is compatible with joint and traditional advertising, and enhances generalizability. Extensive synthetic and industrial experiments demonstrate that JEANet outperforms baselines in revenue while preserving user experience, enabling a platform-user win-win outcome. In summary, our contributions are as follows:

{\bfseries A joint ad mechanism with externalities.} Unlike existing joint auction mechanisms focusing on ad list allocation and pricing, we frame hybrid list generation as a uniform problem. Integrating contextual features, we model ads and organic items to co-determine slot. To our knowledge, this is the first mechanism addressing global externalities and IC challenges arising from separate decision making for ads and organic items in joint advertising.

{\bfseries Designing an adaptive extraction module for bids.} Multi-party advertiser in joint auctions necessitate a store-brand joint selling relationship, aligned with shared marketing objectives. Owing to disparities in advertisers' private values for advertisements, both store and brand advertisers exhibit distinct bid distributions, regardless of their participation in joint selling. Leveraging this insight, we design an adaptive extraction module to capture these distributional differences in multi-party advertiser' bids.

{\bfseries An AMD framework with compatibility.} This framework can flexibly auction ad slots to joint advertisers (virtual entities of store-brand joint selling relationships with shared bidding strategies), store advertisers, or brand advertisers through the proposed bid pairs, and does not rely on the relationship graph of the previous model (e.g., bundles). This dual functionality compatibly supports both joint and traditional advertising, addressing the reduced ad candidate set under equivalent supply (a key revenue-reducing factor) and greatly boosting the practical viability of joint advertising.


{\bfseries Numerical experiments on synthetic and industrial dataset.} We conducted multiple experiments on synthetic and industrial dataset. Compared with commonly used and theoretical baselines, the JEANet-generated mechanism outperforms across metrics.

\section{Related Work}
Previous work focused on externalities in traditional advertising and AMD for the joint ad model, respectively. The key points of this work are summarized below:

{\bfseries AMD for Traditional Advertising with Externalities.} The design of advertising mechanisms accounting for externalities has been a prominent research topic in both academia and industry, with studies evolving from local to global models. Nonetheless, local models \cite{liu2021neural,huang2021deep,liao2022cross,wang2019learning,xie2021hierarchical} cannot generate globally optimal hybrid lists. Consequently, recent research has increasingly focused on global externalities \cite{gatti2012truthful,liao2022nma}. The MIAA \cite{li2024deep} explicitly models the interplay between ads and organic items, determining ad allocation via list-wise pCTR prediction and employing an affine maximization auction (AMA) for payment. Yet, AMAs are limited by their reliance on affine-maximizing mechanisms \cite{wang2024gemnet}. TICNet \cite{ma2025context} jointly optimizes ads and organic items under global information, balancing platform revenue and user experience with approximate DSIC and IR constraints. However, it overlooks the impact of data distribution disparities, particularly the lognormal distribution, which yields suboptimal performance compared to uniform or normal distributions. Moreover, existing global models cannot be readily adapted to joint auctions, as they neglect the distributional characteristics of diverse advertiser types (brand ads, store ads, or joint ads) and the nuances of joint selling relationships. To address these challenges, we propose JEANet, a framework tailored to joint advertising that captures externalities and enables advertising mechanism design.

{\bfseries AMD for Joint Advertising.} The classic auction mechanism has been widely adopted in online advertising systems. The Vickrey-Clarke-Groves (VCG) mechanism \cite{clarke1971multipart,groves1973incentives,vickrey1961counterspeculation} maximizes the social welfare of the bidder without relying on prior distributions, ensuring both DSIC and IR. In contrast, the Myerson auction \cite{myerson1981optimal} optimizes revenue under single-parameter settings. The generalized second price (GSP) auction \cite{caragiannis2011efficiency,edelman2007internet,gomes2009bayes,lucier2012revenue} and its variants \cite{lahaie2007revenue,thompson2013revenue} offer computationally efficient and interpretable models, making them among the most prevalent mechanisms in industry. However, Myerson auctions face challenges in justifying mechanism fairness to advertisers, GSPs lack IC, and VCG is relatively complex.

Traditional theory has established optimal mechanisms for single-item auction, yet extending these to multi-slot auctions remains a significant challenge. With the rise of machine learning, AMD has emerged as a new paradigm for multi-slot auction design \cite{sandholm2015automated}, addressing the limitations of traditional approaches. This framework leverages machine learning techniques to derive optimal auction designs \cite{balcan2008reducing,dutting2015payment,lahaie2011kernel}. Dütting et al. \cite{dutting2024optimal} pioneered the formulation of auction problems as constrained mathematical planning tasks, using regret values to quantify violations of DSIC. They proposed RegretNet, the first neural network framework for multi-slot auctions, which achieves high returns while ensuring IR and approximate DSIC compliance. Subsequent research has expanded RegretNet to address diverse constraints and objectives across auction scenarios, including fairness \cite{kuo2020proportionnet}, budget constraints \cite{feng2018deep}, human preferences \cite{peri2021preferencenet}, attention mechanisms \cite{ivanov2022optimal}, anonymous symmetric auctions \cite{rahme2021permutation}, and contextual auctions \cite{duan2022context}. In the domain of joint auctions, Ma et al. \cite{ma2024joint} first proposed a joint advertising framework and developed the JAMA, a VCG-based mechanism for joint auction design. However, due to the suboptimal revenue performance of JAMA, Zhang et al. \cite{zhang2024joint} designed JRegNet, a RegretNet-based model aimed at maximizing revenue in joint auctions. Despite its improvements, JRegNet is based on store-brand bundles and thus cannot well suit both traditional advertising and joint advertising, thereby restricting its generalizability. Theoretically, Aggarwal et al. \cite{aggarwal2024selling} modeled joint advertising as a multi-period decision problem. Nevertheless, these studies overlook the externality challenges inherent in joint auctions. This paper introduces JEANet, an automated neural network architecture that computes joint auction mechanisms accounting for externalities while satisfying IR and approximate DSIC.

\section{Model and Preliminaries} 
In this section, we first outline the basic framework of external and adaptive joint auctions. We then formalize the joint advertising mechanism design as a dual problem comprising both constrained optimization and learning components.

\subsection{External and Adaptive Joint Auction}
On an Internet advertising platform, when a user initiates a search request, the platform returns a hybrid list of organic items and ads. We consider a page that contains $K$ advertisement slots. For each ad slot $k\in[K] = \{1,\cdots,K\}$, we denote the click-through rate (CTR) of the ad slot $k$ as $\alpha_k$. For simplicity, we assume that the CTRs of the ad slots $k$ are in descending order, that is, $1>\alpha_1\geq\cdots\geq\alpha_K\geq0$. Consider that there are $m\in[M] = \{1,\cdots,M\}$ advertisers and $n\in[N] = \{1,\cdots,N\}$ organic items competing for these $K$ ad slots.

In an external and adaptive joint ad model, store, brand, and joint advertisers can all participate in the auction for each ad slot. In this case, the joint advertiser is a virtual advertiser built on a joint selling relationship, where the store and brand advertisers in the joint selling relationship bid together. Thus, for advertisers $m\in\{M_{store}, M_{brand}, M_{joint}\}$ belonging to different types, the corresponding bid profiles are bid pairs $\mathbf{b}\in\{\mathbf{b_{store}}=(b_{store},0),\mathbf{b_{brand}}=(0,b_{brand}),\mathbf{b_{joint}}=(b_{store},b_{brand})\}$. Every advertiser $i \in [m]$ possesses an ad context denoted by $c_i^{\text{ad}} \in \mathcal{C}^{\text{ad}} \subset \mathbb{R}^{d_c}$. Similarly, each organic item $j \in [n]$ is associated with an organic item context represented as $c_j^{\text{na}} \in \mathcal{C^{\text{na}}} \subset \mathbb{R}^{d_c}$. Specifically, $d_c$ represents the dimension of ad context and organic item context. Then we denote $\mathbf{c^{ad}}=(c_1^{ad},\cdots,c_m^{\text{ad}})$ as the ad contexts and $\mathbf{c^{na}}=(c_1^{na},\cdots,c_n^{\text{na}})$ as the organic items contexts. $\mathbf{c^{ad}}$ and $\mathbf{c^{na}}$ are sampled from publicly known distributions $\mathcal{D}_c^{ad}$ and $\mathcal{D}_c^{na}$.

For each advertiser $i$, the private value for each click by a user is denoted by $v_i$. Assume that the value $v_i$ is sampled from a distribution $\mathcal{F}_{v_i|c_i^{ad},\alpha}$, which depends on the ad context and other slot contexts, over the domain set $ \mathcal{V}_i$. Let $ \mathbf{v}=\{v_1,\cdots,v_m\}$ represent the value profile and $\mathcal{V}=\mathcal{V}_1\times\mathcal{V}_2\times\cdots\times\mathcal{V}_m$ stand for the joint value domain set. Besides, we use $ \mathbf{v}_{-i}$ to express the value profile of all advertisers except $i$, and $ \mathcal{V}_{-i}$ to represent the joint value domain set except $i$. Since advertisers may submit their bids strategically, we utilize $b_i$ as the bid of advertiser $i$, which depends on the private value $v_i$. Similarly, we define $\mathbf{b}$ and $\mathbf{b}_{-i}$ as related bid profiles accordingly.

Additionally, we assume that displaying ads or organic items to a user has a quantifiable impact on their subsequent behavior. Specifically, when users receive a hybrid list containing both ads and organic items from the platform, they form different impressions of each item. These impressions are influenced by various factors, including the perceived value, creativity, and other attributes of the items, as well as the user's preferences. We model this impression as the user experience, which reflects the user's intention to click and propensity to purchase. For each advertiser $i\in [m]$ and organic item $j\in [n]$, we use $ue_i$ and $ue_j$ to denote the expected user experience, respectively. The user experience profile, denoted by $\mathbf{ue} = \{ue_1,\dots,ue_m,ue_{m + 1},\cdots,ue_{m + n}\}$, encompasses both ads and organic items. Specifically, the user experience $ue_i$ of advertiser $i$ is sampled from a distribution $\mathcal{F}_{ue_i|c_i^{ad},\alpha}^{ad}$, and $ue_j$ of organic item $j$ is sampled from another distribution $ \mathcal{F}_{ue_j|c_j^{na},\alpha}^{na}$. Similar to the definition of the joint value distribution, let $\mathcal{UE} = \mathcal{UE}_1\times\mathcal{UE}_2\times\cdots\times\mathcal{UE}_m\times\mathcal{UE}_{m + 1}\times\cdots\times\mathcal{UE}_{m + n}$ be the joint user experience domain set of ads and organic items.

Having introduced the background, we now formally introduce the joint advertisement auction with compatibility and externalities, i.e., $\mathcal{M} = (\mathbf{a}, \mathbf{p})$, which consists of an allocation rule $\mathbf{a}(\mathbf{b}, \mathbf{ue}, \mathbf{c^{ad}}, \mathbf{c^{na}}, \mathbf{\alpha}) = \{a_i(\mathbf{b}, \mathbf{ue}, \mathbf{c^{ad}}, \mathbf{c^{na}}, \mathbf{\alpha}\}_{i \in [m] \cup [n]}$ and a payment scheme $ \mathbf{p}(\mathbf{b}, \mathbf{ue}, \mathbf{c^{ad}}, \mathbf{c^{na}}, \mathbf{\alpha}) = \{p_i(\mathbf{b}, \mathbf{ue}, \mathbf{c^{ad}}, \mathbf{c^{na}}, \mathbf{\alpha})\}_{i \in M}$. To be specific, $a_i(\mathbf{b}, \mathbf{ue}, \mathbf{c^{ad}}, \mathbf{c^{na}}, \mathbf{\alpha}): \mathcal{V} \times \mathcal{UE} \times \mathcal{C^{\text{ad}}_\text{m}} \times \mathcal{C^{\text{na}}_\text{n}} \times \mathbf{\alpha}\to [0, 1]$ stands for the expected CTR that ad (or organic item) $i$ can derive from being shown to the user. That is, $a_i = \sum_{k = 1}^{K} a_{ik}(\mathbf{b}, \mathbf{ue}, \mathbf{c^{ad}}, \mathbf{c^{na}})\alpha_k$, where $a_{ik}(\mathbf{b}, \mathbf{ue}, \mathbf{c^{ad}}, \mathbf{c^{na}}) \in \{0, 1\}$ indicates whether item $i$ is allocated to slot $k$ or not. As for the payment rule, $p_i(\mathbf{b}, \mathbf{ue}, \mathbf{c^{ad}}, \mathbf{c^{na}}, \mathbf{\alpha})$ specifies the payment that advertiser $i$ must pay per click.

In an external and adaptive joint ad auction setup with compatibility and externality constraints, where multi-party advertiser (stores, brands, and joint ads) are included, each advertiser can occupy at most one ad slot, and each ad slot is assigned to exactly one item (either an ad or an organic item). The allocation rule constraints, referred to as Feasibility \cite{dutting2024optimal}, are specified as follows:

\begin{equation*}
\begin{cases} 
\begin{aligned}
    &\sum_{k \in [K]} a_{ik}(\mathbf{b}, \mathbf{ue}, \mathbf{c^{ad}}, \mathbf{c^{na}}) \leq 1,  &\forall i \in [m] \cup [n], \\
    &\sum_{i \in [m] \cup [n]} a_{ik}(\mathbf{b}, \mathbf{ue}, \mathbf{c^{ad}}, \mathbf{c^{na}}) = 1,  &\forall k \in [K], \\
    &a_{ik}(\mathbf{b}, \mathbf{ue}, \mathbf{c^{ad}}, \mathbf{c^{na}}) \in \{0, 1\}, &\forall i \in [m] \cup [n],\ k \in [K].
\end{aligned}
\end{cases} 
\end{equation*}
Under the setting of an external and adaptive joint ad auction, each advertiser pursues optimizing the quasi-linear utility, defined as:
\begin{equation*}
    u_i \cdot (v_i; \mathbf{b}, \mathbf{ue}, \mathbf{c^{ad}}, \mathbf{c^{na}}, \mathbf{\alpha}) = v_i \cdot a_i(\mathbf{b}, \mathbf{ue}, \mathbf{c^{ad}}, \mathbf{c^{na}}) - p_i \cdot (\mathbf{b}, \mathbf{ue}, \mathbf{c^{ad}}, \mathbf{c^{na}}, \mathbf{\alpha})
\end{equation*}
for all $i\in [m]$, $v_i \in \mathcal{V}_i$, $\mathbf{b} \in \mathcal{V}$, $\mathbf{ue} \in \mathcal{UE}$, $\mathbf{c^{ad}} \in \mathcal{C^{\text{ad}}_\text{m}}$, $\mathbf{c^{na}} \in \mathcal{C^{\text{na}}_\text{n}}$.

In auction mechanism design, DSIC and IR are two fundamental economic properties: DSIC ensures that the optimal strategy for bidders is to truthfully reveal their private values, while IR ensures that bidders always derive non-negative utility.

\begin{definition}[Dominant Strategy Incentive Compatibility]
An external and adaptive joint ad auction $(\mathbf{a}, \mathbf{p})$ is dominant strategy incentive compatible, if for any advertiser, the utility is maximized by truthful reporting regardless of others’ reports. Formally, it holds that for all $i \in [m]$, $v_i \in \mathcal{V}_i$, $b_i' \in \mathcal{V}_i$, $\mathbf{ue} \in \mathcal{UE}$, $\mathbf{c^{ad}} \in \mathcal{C^{\text{ad}}_\text{m}}$, $\mathbf{c^{na}} \in \mathcal{C^{\text{na}}_\text{n}}$,
\begin{equation*}
    u_i(v_i; (v_i, \mathbf{b}_{-i}), \mathbf{ue}, \mathbf{c^{ad}}, \mathbf{c^{na}}, \mathbf{\alpha}) \geq u_i(v_i; (b_i', \mathbf{b}_{-i}), \mathbf{ue}, \mathbf{c^{ad}}, \mathbf{c^{na}}, \mathbf{\alpha}).
\end{equation*}
\end{definition}

\begin{definition}[Individual Rationality]
An external and adaptive joint ad auction is individual rational, if for any advertiser, the utility will be non-negative when they bid truthfully. Formally, it holds that, for all $i \in [m]$, $v_i \in \mathcal{V}_i$, $b_i' \in \mathcal{V}_i$, $\mathbf{ue} \in \mathcal{UE}$, $\mathbf{c^{ad}} \in \mathcal{C^{\text{ad}}_\text{m}}$, $\mathbf{c^{na}} \in \mathcal{C^{\text{na}}_\text{n}}$,
\begin{equation*}
    u_i(v_i; (v_i, \mathbf{b}_{-i}), \mathbf{ue}, \mathbf{c^{ad}}, \mathbf{c^{na}}, \mathbf{\alpha}) \geq 0.
\end{equation*}
\end{definition}

In an external and adaptive joint ad auction satisfying DSIC and IR, advertisers will truthfully bid. Let $\mathcal{F}_{\mathbf{v}, \mathbf{ue}, \mathbf{c^{ad}}, \mathbf{c^{na}}, \mathbf{\alpha}}$ denote the joint distribution of the value profile $\mathbf{v}$, user experience $\mathbf{ue}$, ad contexts $ \mathbf{c^{ad}}$, organic contexts $\mathbf{c^{na}}$ and ad slot CTR $\mathbf{\alpha}$. The platform focuses on two core metrics: revenue and user experience. The platform’s expected revenue is defined as:
\begin{equation*}
\begin{aligned}
 \text{Rev} := \mathbb{E}_{(\mathbf{v}, \mathbf{ue}, \mathbf{c^{ad}}, \mathbf{c^{na}}, \mathbf{\alpha}) \sim \mathcal{F}}\left[ \sum_{i \in [m]} p_i(\mathbf{v}, \mathbf{ue}, \mathbf{c^{ad}}, \mathbf{c^{na}}, \mathbf{\alpha}) \right].
\end{aligned}
\end{equation*}
while the expected user experience is defined as:
\begin{equation*}
\begin{aligned}
 \text{UE} := \mathbb{E}_{(\mathbf{v}, \mathbf{ue}, \mathbf{c^{ad}}, \mathbf{c^{na}}, \mathbf{\alpha}) \sim \mathcal{F}}\left[ 
    \sum_{i \in [m] \cup [n]} ue_i \cdot a_i(\mathbf{v}, \mathbf{ue}, \mathbf{c^{ad}}, \mathbf{c^{na}}, \mathbf{\alpha})
\right].
\end{aligned}
\end{equation*}

We aim to design an external and adaptive joint ad auction that optimizes the trade-off between revenue and user experience while preserving DSIC and IR. To balance revenue and user experience, we introduce a hyperparameter 
$\gamma > 0$. Formally, our problem can be formulated as a constrained optimization problem:
\begin{equation*}
\begin{aligned}
    \max_{(\mathbf{a}, \mathbf{p})} \quad &\mathbb{E}_{(\mathbf{v}, \mathbf{ue}, \mathbf{c^{ad}}, \mathbf{c^{na}}, \mathbf{\alpha}) \sim \mathcal{F}_{\mathbf{v}, \mathbf{ue}, \mathbf{c^{ad}}, \mathbf{c^{na}}, \mathbf{\alpha}}}\Bigg[ 
        \sum_{i \in [m]} p_i(\mathbf{v}, \mathbf{ue}, \mathbf{c^{ad}}, \mathbf{c^{na}}, \mathbf{\alpha}) \\
        &\quad + \gamma \cdot \sum_{i \in [m] \cup [n]} ue_i \cdot a_i(\mathbf{v}, \mathbf{ue}, \mathbf{c^{ad}}, \mathbf{c^{na}}, \mathbf{\alpha}) 
    \Bigg] \\
    \text{s.t.} \quad &\text{DSIC, IR, Feasibility.}
\end{aligned}
\end{equation*}

\begin{figure*}
  \centering
  \includegraphics[width=16cm]{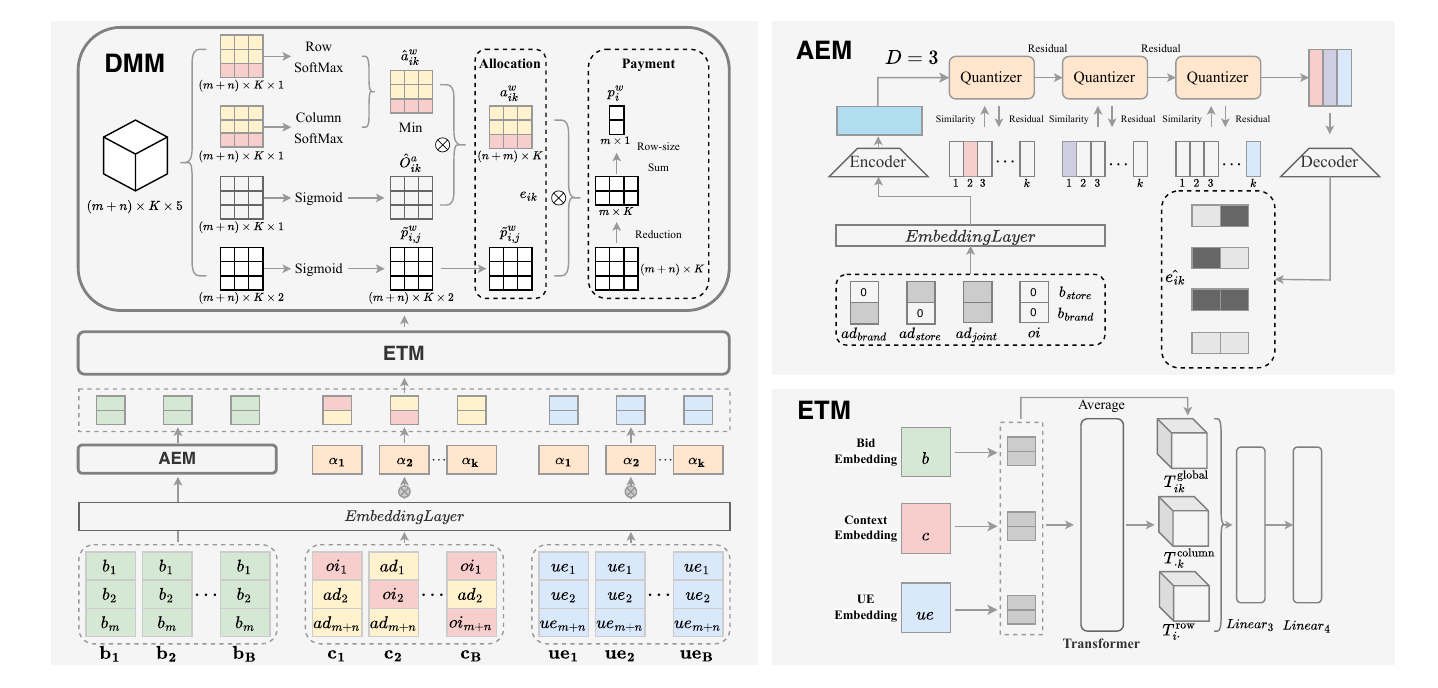}
  \caption{The architecture of JEANet. JEANet consists of three modules: Adaptive Extraction Module (AEM), Externality Transformer Module (ETM) and Deep Mechanism Module (DMM).}
  \label{fig:model}
  \Description{The architecture of JEANet. JEANet consists of three modules: Adaptive Extraction Module (AEM), Externality Transformer Module (ETM) and Deep Mechanism Module (DMM).}
\end{figure*}

\subsection{External and Adaptive Joint Auction Design as a Learning Problem}
We now formulate the problem of designing an external and adaptive joint ad auction as a learning problem. To ensure DSIC, we introduce the metric of ex-post regret. For advertiser $i$, the regret is defined as the maximum utility they could gain by misreporting their valuation while holding other advertisers' bids fixed, i.e.,

\begin{equation*}
\begin{aligned}
    \text{rgt}_i(\mathbf{v}, \mathbf{ue}, \mathbf{c}^\text{ad}, \mathbf{c}^\text{na}, \mathbf{\alpha}) 
    &= \max_{b_i' \in \mathcal{V}_i} \Big[ 
        u_i\big(v_i; (b_i', \mathbf{b}_{-i}), \mathbf{ue}, \mathbf{c}^\text{ad}, \mathbf{c}^\text{na}, \mathbf{\alpha}\big) \\ 
    &\quad - u_i\big(v_i; (v_i, \mathbf{b}_{-i}), \mathbf{ue}, \mathbf{c}^\text{ad}, \mathbf{c}^\text{na}, \mathbf{\alpha}\big) 
    \Big].
\end{aligned}
\end{equation*}

We can draw the conclusion that an external and adaptive joint ad auction achieves DSIC if and only if its regret equals 0.
Subsequently, we reformulate the problem of designing the joint ad auction with such constraints as a constrained optimization problem:
\begin{equation*}
\begin{aligned}
    \min_{(\mathbf{a}, \mathbf{p}) \in \mathcal{M}} \quad &- \mathbb{E}_{(\mathbf{v}, \mathbf{ue}, \mathbf{c^{ad}}, \mathbf{c^{na}}, \mathbf{\alpha}) \sim \mathcal{F}}\Bigg[ 
        \sum_{i \in [m]} p_i(\mathbf{v}, \mathbf{ue}, \mathbf{c^{ad}}, \mathbf{c^{na}}, \mathbf{\alpha}) \\
        &\quad + \gamma \cdot \sum_{i \in [m] \cup [n]} ue_i \cdot a_i(\mathbf{v}, \mathbf{ue}, \mathbf{c^{ad}}, \mathbf{c^{na}}, \mathbf{\alpha}) 
    \Bigg] \\
    \text{s.t.} \quad & \mathbb{E}_{(\mathbf{v}, \mathbf{ue}, \mathbf{c^{ad}}, \mathbf{c^{na}}, \mathbf{\alpha}) \sim \mathcal{F}}\left[ 
        \sum_{i \in [m]} \text{rgt}_i(\mathbf{v}, \mathbf{ue}, \mathbf{c^{ad}}, \mathbf{c^{na}}, \mathbf{\alpha}) 
    \right] = 0.
\end{aligned}
\end{equation*}
where $\mathcal{M}$ is the set of all external and adaptive joint ad auctions that satisfy IR and Feasibility. 
However, the problem remains intractable due to the high complexity of the constraints. 
To solve this problem, we parameterize the auction mechanism as 
$\mathcal{M}^w(\mathbf{a}^w, \mathbf{p}^w) \subseteq \mathcal{M}(\mathbf{a}, \mathbf{p})$, 
where $w \in \mathbb{R}^{d_w}$ (with dimension $d_w$) are the parameters. Then we aim to find an external and adaptive joint ad auction $\mathcal{M}^w$ that minimizes the negated objective:
\begin{equation*}
    - \mathbb{E}_{\mathbf{} \sim \mathcal{F}_{\mathbf{}}}\Bigg[ 
        \sum_{i \in [m]} p_i^w 
        + \gamma \cdot \sum_{i \in [m] \cup [n]} ue_i \cdot a_i^w 
    \Bigg].
\end{equation*}
while ensuring DSIC, IR and Feasibility constraints through optimizing the parameters.

Given a sample $\mathcal{L}$, containing $L$ samples of $(\mathbf{v}, \mathbf{ue}, \mathbf{c^{ad}}, \mathbf{c^{na}}, \mathbf{\alpha})$ drawn from the distribution $\mathcal{F}_{\mathbf{v}, \mathbf{ue}, \mathbf{c^{ad}}, \mathbf{c^{na}}, \mathbf{\alpha}}$, we can estimate the ex-post regret under $\mathcal{M}^w(\mathbf{a}^w, \mathbf{p}^w)$ as:
\begin{equation*}
    \widehat{\text{rgt}}_i(w) = \frac{1}{L} \sum_{\ell=1}^{L} \text{rgt}_i\big(\mathbf{v}_{(\ell)}, \mathbf{ue}_{(\ell)}, \mathbf{c}^\text{ad}_{(\ell)}, \mathbf{c}^\text{na}_{(\ell)}, \mathbf{\alpha}_{(\ell)}\big).
\end{equation*}

To sum up, we can formulate our constrained optimization problem as a learning problem as follows:
\begin{equation*}
\begin{aligned}
    \min_{w \in \mathbb{R}^{d_w}} \quad &- \frac{1}{L} \sum_{\ell=1}^{L} \Bigg[ 
        \sum_{i \in [m]} p_i^w(\mathbf{v}_{(\ell)}, \mathbf{ue}_{(\ell)}, \mathbf{c}^\text{ad}_{(\ell)}, \mathbf{c}^\text{na}_{(\ell)}, \mathbf{\alpha}_{(\ell)}) \\
        &+ \gamma \cdot \sum_{i \in [m] \cup [n]} \alpha_i \cdot a_i^w(\mathbf{v}_{(\ell)}, \mathbf{ue}_{(\ell)}, \mathbf{c}^\text{ad}_{(\ell)}, \mathbf{c}^\text{na}_{(\ell)}, \mathbf{\alpha}_{(\ell)}) 
    \Bigg] \\
    \text{s.t.} \quad & \widehat{rgt}_i(w) = 0, \quad i = 1, \cdots, m.
\end{aligned}
\end{equation*}

Additionally, we guarantee that our auction mechanism satisfies IR through the structure of our architecture.

\section{JEANet}
In this section, we formalize the JEANet to design joint advertising auction mechanisms.


\subsection{The JEANet Architecture}
As illustrated in \autoref{fig:model}, JEANet comprises three components: the Adaptive Extraction Module (AEM), the Externality Transformer Module (ETM), and the Deep Mechanism Module (DMM). Specifically, AEM takes embedded ads and organic items, preprocesses, combines them, and generates a feature matrix via stacked Residual-Quantized (RQ) layers \cite{lee2022autoregressive}—a self-supervised clustering method proven effective in identifying latent classes without prior knowledge. The clustering effect is independent of codebook size, effectively resolving the conflict between the VQ layer’s clustering performance and codebook size. This matrix is then fed into the ETM, which uses a Transformer-based encoder to process it, modeling implicit relationships between ads and organic items. The DMM takes the ETM output and outputs the allocation of ads and organic items (satisfying the DSIC, IR, and Feasibility constraints) as well as the advertisers’ payments.
\subsection{Adaptive Extraction Module}
First of all, we apply embedding layers to obtain a representation $t_i \in \mathbb{R}^{d_c'}$ for each ad context $c^{ad}_i$ , $t_j \in \mathbb{R}^{d_c'}$ for each organic item context $c^{na}_j$. Afterward, as mentioned before, each advertiser $i$ has a bid profile and a user experience profile, while each organic item $j$ is only equipped with a user experience profile. We assume that organic item $j$ can also submit a bid to the platform, with $\mathbf{b_{na}}=[0, 0]$. Unlike advertisers, every organic item $j$ cannot intentionally misreport its bid. Under the above assumption, we map bid profile embeddings and user experience profile embeddings into $\mathbf{e} = (e_{ik})_{i \in [m] \cup [n], k \in [K]}$ and $\mathbf{z} = (z_{jk})_{j \in [m] \cup [n], k \in [K]}$, where $e_{ik}$ and $z_{jk} \in \mathbb{R}^{d_z'}$ are defined as:
\begin{equation*}
\begin{cases} 
    z_{jk} = Emb(ue_j) \cdot \alpha_k, & \forall j \in [m] \cup [n],\ \forall k \in [K], \\
    e_{ik} = b_i \cdot \alpha_k, & \forall i \in [m],\ \forall k \in [K], \\
    e_{ik} = 0, & \forall i \in [n],\ \forall k \in [K].
\end{cases}
\end{equation*}

We use stackable RQ layers to discretize bid profiles $\mathbf{e}$. Let codebook $\mathcal{C}$ be a finite set $\{(u, \mathbf{t}(u))\}_{u \in [U]}$ of codes $u$ and their embeddings $\mathbf{t}(u) \in \mathbb{R}^{d_c'}$, where $U$ is the size and $d_c'$ is the embedding dimension. For bid profiles $\mathbf{e}$, $\mathcal{Q}(\mathbf{e}; \mathcal{C})$ denotes its VQ, the code with the embedding nearest to $\mathbf{e}$:
\begin{equation*}
    \mathcal{Q}(\mathbf{e}; \mathcal{C}) = \underset{u \in [U]}{\arg\min} \|\mathbf{e} - \mathbf{t}(u)\|_2^2.
\end{equation*}

Obviously, the expressiveness of quantization is limited by the codebook size, but increasing it harms efficiency. Instead, we use RQ to discretize bid profiles $\mathbf{e}$.
\begin{equation*}
    \mathcal{RQ}(\mathbf{e}; \mathcal{C}, D) = (t_1, \cdots, t_D) \in [U]^D.
\end{equation*}
where $\mathcal{C}$ is the codebook with size $|\mathcal{C}| = U$, and $u_d$ is the code of $\mathbf{e}$ at depth $d$. Starting with the 0-th residual $r_0=\mathbf{e}$, RQ recursively computes $u_d$, which is the code of the residual $r_{d-1}$, and the next residual $r_d$ as
\begin{equation*}
\begin{aligned}
    u_d &= \mathcal{Q}(\mathbf{r}_{d - 1}; \mathcal{C}), \\
    \mathbf{r}_d &= \mathbf{r}_{d - 1} - \mathbf{t}(u_d).
\end{aligned}
\end{equation*}
for $d = 1, \cdots, D$. In addition, we define $\hat{\mathbf{e}}^{(d)} = \sum_{i = 1}^{d} \mathbf{t}(u_i)$ as the partial sum of up to $d$ code embeddings, and $\hat{\mathbf{e}} := \hat{\mathbf{e}}^{(D)}$ as the quantized vector of $\mathbf{e}$. Finally, the decoder $G$ reconstructs the input from $\hat{\mathbf{e}}$ as $\hat{\mathbf{R}} = G(\hat{\mathbf{e}}).$ With the above representation, we obtain the ad-organic items overall representation $R = (I_{i,k})_{i \in [m] \cup [n], k \in [K]}$:
\begin{equation*}
\begin{aligned}
    I_{i,k} = [\hat{e_{ik}}||z_{ik}||t_i] \in \mathbb{R}^{2 + d_c' + d_z'}.
\end{aligned}
\end{equation*}
where “||” stands for concatenation.

In order to enhance the efficiency of JEANet, we utilize two linear layers with a ReLU activation to reduce the last dimension of $I$ from $2 + d_c' + d_z'$ to $d' - 2$, i.e.,
\begin{equation*}
\begin{aligned}
    I' = \text{Linear}_2(\text{ReLU}(\text{Linear}_1(I))) \in \mathbb{R}^{(m + n) \times K \times (d' - 2)}.
\end{aligned}
\end{equation*}

After concatenating $I'$ with transferred bid profiles and user experience profiles, we can obtain the output of the input layer:
\begin{equation*}
\begin{aligned}
    T = [I'||\mathbf{e}||\mathbf{z}] \in \mathbb{R}^{(m + n) \times K \times d'}.
\end{aligned}
\end{equation*}
in which $T_{ik} \in T$ provide the joint representation of the bid, the user experience and the corresponding context for the ad (or the organic item) in ad slot $k$.

\subsection{Externality Transformer Module}
Using AEM's feature representation $T \in \mathbb{R}^{(m + n) \times K \times d'}$, we focus on inner relationships among ads, organic items and slots, using a Transformer encoder to capture their complex interactions. To be specific, we employ a Transformer-based encoder \cite{ma2025context} to model the interactions between ads (or organic items) $i$ and different slots by operating row-wisely on $T$:
\begin{equation*}
\begin{aligned}
    T_{i\cdot}^{\text{row}} = \text{Transformer}(T_{i\cdot}) \in \mathbb{R}^{(m + n) \times d_h}, \quad \forall i \in [m] \cup [n].
\end{aligned}
\end{equation*}
where $d_h$ denotes the encoder MLP's hidden layer dimension. Similarly, a column-wise encoder is applied to $T$ 's $k$-th column to model interactions between slot $k$ and all ads and organic items:

\begin{equation*}
\begin{aligned}
    T_{\cdot k}^{\text{column}} = \text{Transformer}(T_{\cdot k}) \in \mathbb{R}^{K \times d_h}, \quad \forall k \in [K]. 
\end{aligned}
\end{equation*}
In addition, we conduct averaging on $T$ to obtain the global representation.
Then we concatenate $T^{\text{row}}$, $T^{\text{column}}$ and $T^{\text{global}}$ together along the last dimension and obtain $T' = (T_{ik}')_{i \in [m] \cup [n], k \in [K]}$.

Afterward, we apply two linear layers with a ReLU activation to reduce the last dimension of $T'$ from $2 \cdot d_h + d'$ to $d_{\text{out}}$ , i.e.,
\begin{equation*}
\begin{aligned}
    O = \text{Linear}_4(\text{ReLU}(\text{Linear}_3(T'))) \in \mathbb{R}^{(m + n) \times K \times d_{\text{out}}}.
\end{aligned}
\end{equation*}
where $O$ is the output of the ETM.

\subsection{Deep Mechanism Module}
With the DMM, setting $d_{\text{out}} = 5$ gives the representation $O = (O^r, O^c, O^a, O^{p_1}, O^{p_2}) \in \mathbb{R}^{(m + n) \times K \times 5}$, which is used to compute allocation results and winning advertisers' payments. Next, we compute the allocation output $a_{ik}^w$ (the probability of ad or organic item $i$ being allocated to slot $k$). To ensure Feasibility, we apply row-wise softmax to $O^r$ to obtain $\hat{O}^r$, which ensures that  each ad (or organic item) $i$ will obtain no more than one slot. Similarly, we apply column-wise softmax to $O^c$ to obtain $\hat{O}^c$, which guarantees that no slot is over-allocated. Formally, $\hat{O}^r$ and $\hat{O}^c$ are defined as:
\begin{equation*}
\begin{cases}
    \hat{O}_{i\cdot}^r = \text{Softmax}(O_{i\cdot}^r), & \forall i \in [m]        \cup [n], \\ \hat{O}_{\cdot k}^c = \text{Softmax}(O_{\cdot k}^c), & \forall k \in [K].
\end{cases}
\end{equation*}
Then we obtain the original allocation probability by:
\begin{equation*}
\begin{aligned}
    \hat{a}_{ik}^w = \min\{\hat{O}_{ik}^r, \hat{O}_{ik}^c\}, \quad \forall i \in [m] \cup [n], \forall k \in [K].
\end{aligned}
\end{equation*}
Afterward, some ads and organic items may fail to get slots but retain part of the original allocation. To reduce their impact on the final results, we use $O^a$ to adjust the original allocation probabilities. Essentially, we apply the sigmoid activation to $O^a$ to obtain $\hat{O}^a$, then multiply it by $\hat{a}_{ik}^w$ to get the final allocation $a_{ik}^w$:
\begin{equation*}
\begin{cases}
    \hat{O}_{ik}^a = \text{Sigmoid}(O_{ik}^a), \forall i \in [m] \cup [n],\ \forall k \in [K], \\ 
    a_{ik}^w = \hat{a}_{ik}^w \cdot \hat{O}_{ik}^a.
\end{cases}
\end{equation*}

The deterministic allocation with the maximum probability obtained by referring to Birkhoff \cite{birkhoff1946tres,dutting2024optimal}. Then we proceed to calculate the payment. Initially, we compute the auxiliary payment scalar based on $O^{p_j} \left(j \in \{1, 2\}\right)$ using the sigmoid activation:
\begin{equation*}
\begin{aligned}
    \tilde{p}_{i,j}^{w} = \text{Sigmoid}(\frac{1}{K} \sum_{k \in K} O^{p_j}), \quad \forall i \in [m].
\end{aligned}
\end{equation*}

Given the auxiliary payment scalar and final allocation result, we then obtain the final payment result as:
\begin{equation*}
\begin{aligned}
    p_{i,j}^{w} = \tilde{p}_{i,j}^{w} \cdot \left[ \sum_{k \in [K]} a_{ik}^w \cdot e_{ik} \right].
\end{aligned}
\end{equation*}

Since $\tilde{p}_{i,j}^{w} \in (0, 1)$, the quasi-linear utility of advertiser $i$, denoted as $u_i^w = \sum_{j \in \{1, 2\}} \left( \sum_{k \in [K]} a_{ik}^w \cdot e_{ik} - p_{i,j}^{w} \right)$, remains non-negative, thereby ensuring the IR of our mechanism.

\subsection{Training Procedure}
As for the training process of JEANet, the overall $loss$ consists of two parts: $loss_{AEM}$ and $loss_{DMM}$. The two losses are optimized alternately. We first use gradient descent with respect to the $loss_{AEM} = loss_{\text{recon}} + \beta \cdot loss_{\text{commit}}$ \cite{lee2022autoregressive}, where $\beta > 0$ is a multiplicative factor. The reconstruction loss $loss_{\text{recon}}$ and the commitment loss $loss_{\text{commit}}$ are defined as:
\begin{equation*}
\begin{aligned}
    loss_{\text{recon}} &= \|\mathbf{R} - \hat{\mathbf{R}}\|_2^2, \\
    loss_{\text{commit}} &= \sum_{d = 1}^{D} \left\|\mathbf{e} - \text{sg} \left[ \hat{\mathbf{e}}^{(d)} \right] \right\|_2^2.
\end{aligned}
\end{equation*}
where $\text{sg}[\cdot]$ is the stop-gradient operator, and the straight through estimator is used for the backpropagation through the AEM.

After optimizing for several rounds, the AEM parameters are fixed; then the augmented Lagrangian method is used to optimizes the following formula for $loss_{DMM}$. The corresponding formula is:

\begin{equation*}
\begin{aligned}
    C_{\rho}(w; \lambda) =& - \frac{1}{L} \sum_{\ell = 1}^{L} \Bigg[ \sum_{i \in [m]} p_i^w(\mathbf{v}_{(\ell)}, \mathbf{ue}_{(\ell)}, \mathbf{c}^{ad}_{(\ell)}, \mathbf{c}^{na}_{(\ell)}, \mathbf{\alpha}^{(\ell)}) \\
    & + \gamma \cdot \sum_{j \in [m] \cup [n]} ue_j \cdot a_j^w(\mathbf{v}_{(\ell)}, \mathbf{ue}_{(\ell)}, \mathbf{c}^{ad}_{(\ell)}, \mathbf{c}^{na}_{(\ell)}, \mathbf{\alpha}_{(\ell)}) \Bigg] \\
    & + \sum_{i = 1}^{m} \lambda_i \cdot \widehat{\text{rgt}}_i(w) + \frac{\rho}{2} \sum_{i = 1}^{m} (\widehat{\text{rgt}}_i(w))^2.
\end{aligned}
\end{equation*}

In optimization, we first calculate the optimal misreports and then alternately update model parameters and Lagrange multipliers.

\section{Experiments}
In this section, we conduct a series of experiments on the synthetic dataset and industrial dataset to validate the superiority of JEANet and aim to answer the following research questions:

\begin{itemize}[leftmargin=*]
\item \textbf{RQ1}: How does our JEANet perform compared to the state-of-the-art joint advertising models?
\end{itemize}

\begin{itemize}[leftmargin=*]
\item \textbf{RQ2}: How does JEANet perform in real-world industrial advertising scenarios?
\end{itemize}

\begin{itemize}[leftmargin=*]
\item \textbf{RQ3}: Why does the mechanism computed using JEANet satisfy approximate DSIC rather than exact IC?
\end{itemize}

\begin{itemize}[leftmargin=*]
\item \textbf{RQ4}: What is the impact of designs (e.g. AEM, ETM) on the performance of JEANet?
\end{itemize}

\subsection{Experimental Settings}
\subsubsection{Dataset}
We offer empirical evidence for the effectiveness of our mechanisms on the synthetic dataset and the industrial dataset, with detailed descriptions provided below:

\begin{itemize}[leftmargin=*]
\item \textbf{Synthetic Dataset}. Synthetic data are generated under various settings, each with 100,000 training and 10,000 testing profiles. Contexts have two feature types: continuous (sampled from multi-dimensional uniform distributions) and discrete (with preset finite types). Advertiser bid profiles and user experience profiles for ads and organic items are generated from their contexts. Two settings are used:

\begin{itemize}[leftmargin=*]
\item \textbf{Random number of items}. We set the total number of ads ($num_{ad}$) and organic items ($num_{na}$) to 20. For each sample, the $num_{ad}$ ($3 \leq num_{ad} \leq 10$) is random, including store ads ($num_{store} \geq 1$), brand ads ($num_{brand} \geq 1$), and joint ads ($num_{joint} \geq 1$), such that their sum $\leq 10$. We then compute the $num_{na} = 20-num_{ad}$. For auctions with 4, 5, or 6 slots, the CTR is set to $\alpha = (0.7,0.6,0.5,0.4)$, $\alpha = (0.7,0.6,0.5,0.4,0.3)$ or $\alpha = (0.7,0.6, 0.5,0.4,0.3,0.2)$, respectively. The bid profiles of advertisers are independently sampled from $U[0,1]$, while user experiences for ads and organic items are independently sampled from $U[0,0.5]$ and $U[0.5,1]$, respectively.
\end{itemize}

\begin{itemize}[leftmargin=*]
\item \textbf{Fixed number of items}. We fix the number of ads and organic items separately, with the following settings:
\begin{itemize}[leftmargin=*]
\item \textbf{(A)}. 4 ads, 6 organic items, and 3 slots with CTR $\alpha = (0.5,0.3, \break 0.2)$. The raw bid of each advertiser is independently drawn from $U[0,1]$. The raw user experience is independently drawn from $U[0,0.5]$ and $U[0.5,1]$.
\end{itemize}

\begin{itemize}[leftmargin=*]
\item \textbf{(B)}. 5 ads, 5 organic items, and 3 slots with CTR $\alpha = (0.5,0.3, \break 0.2)$. The context information, raw profiles, and corresponding transferred profiles are drawn similarly to Setting A.
\end{itemize}

\begin{itemize}[leftmargin=*]
\item \textbf{(C)}. 6 ads, 4 organic items, and 3 slots with CTR $\alpha = (0.5,0.3, \break 0.2)$. The context information, raw profiles, and corresponding transferred profiles are drawn similarly to Setting A.
\end{itemize}

\end{itemize}
\end{itemize}

\begin{table*}
\centering
  \caption{Experimental results for the random number of items in synthetic data. Each VCG metric is set to 1, with $score$ calculated as 1.5, and each column is normalized. The regret of mechanisms generated by Revised JRegNet, Revised TICNet, and JEANet are < 0.001. The best performance is highlighted in bold. ``$\dagger$'' indicates a statistically significant improvement in a paired $t$-test at $p$ < 0.05.}
  \label{tab:first table}
  \begin{tabular}{ccccccccccccc}
    \toprule
    \multirow{2}{*}{Method} & \multicolumn{4}{c}{slot=4} & \multicolumn{4}{c}{slot=5} & \multicolumn{4}{c}{slot=6} \\
    & SW & Rev & UE & Score & SW & Rev & UE & Score & SW & Rev & UE & Score \\
    \midrule
    GSP with Fixed Positions & 0.822 & 1.115 & 1.207 & 1.719 & 0.872 & 1.359 & 1.173 & 1.946 & 0.906 & 1.678 & 1.151 & 2.254 \\
    VCG & \textbf{1.000} & 1.000 & 1.000 & 1.500 & \textbf{1.000} & 1.000 & 1.000 & 1.500 & \textbf{1.000} & 1.000 & 1.000 & 1.500 \\
    IAS & 0.891 & 1.178 & 0.961 & 1.659 & 0.922 & 1.331 & 0.854 & 1.758 & 0.936 & 1.537 & 0.770 & 1.922 \\
    Revised JRegNet & 0.817 & 0.783 & 0.977 & 1.271 & 0.795 & 0.840 & 0.980 & 1.331 & 0.832 & 1.190 & 0.988 & 1.684 \\
    Revised TICNet & 0.825 & 1.209 & 1.316 & 1.867 & 0.770 & 1.297 & \textbf{1.480} & 2.038 & 0.786 & 1.485 & 1.381 & 2.176 \\
    JEANet & 0.864 & \textbf{1.297} & \textbf{1.477} & $\textbf{2.035}^{\dagger}$ & 0.873 & \textbf{1.411} & 1.476 & $\textbf{2.149}^{\dagger}$ & 0.901 & \textbf{1.799} & \textbf{1.389} & $\textbf{2.494}^{\dagger}$ \\
    \bottomrule
  \end{tabular}
\end{table*}

\begin{table*}
\centering
  \caption{Experimental results for the fixed number of items in synthetic data. The notation $m \times n \times k$ represents a setting where $m$ advertisers and $n$ organic items compete for $k$ slots. Each VCG metric is set to 1, with $score$ calculated as 1.5, and each column is normalized. The regret of mechanisms generated by Revised JRegNet, Revised TICNet, and JEANet are < 0.001. The best performance is highlighted in bold. ``$\dagger$'' indicates a statistically significant improvement in a paired $t$-test at $p$ < 0.05.}
  \label{tab:second table}
  \begin{tabular}{ccccccccccccc}
    \toprule
    \multirow{2}{*}{Method} & \multicolumn{4}{c}{A: $4 \times 6 \times 3$} & \multicolumn{4}{c}{B: $5 \times 5 \times 3$} & \multicolumn{4}{c}{C: $6 \times 4 \times 3$} \\
    & SW & Rev & UE & Score & SW & Rev & UE & Score & SW & Rev & UE & Score \\
    \midrule
    GSP with Fixed Positions & 0.787 & 1.326 & 1.253 & 1.953 & 0.763 & 1.094 & 1.221 & 1.705 & 0.751 & 0.980 & 1.196 & 1.578 \\
    VCG & \textbf{1.000} & 1.000 & 1.000 & 1.500 & \textbf{1.000} & 1.000 & 1.000 & 1.500 & \textbf{1.000} & 1.000 & 1.000 & 1.500 \\
    IAS & 0.822 & 1.664 & 1.064 & 2.196 & 0.904 & 1.414 & 1.020 & 1.924 & 0.973 & 1.320 & 1.000 & 1.820 \\
    Revised JRegNet & 0.762 & 1.482 & 1.193 & 2.079 & 0.731 & 1.135 & 1.169 & 1.719 & 0.725 & 0.994 & 1.153 & 1.570 \\
    Revised TICNet & 0.883 & 2.085 & \textbf{1.442} & 2.806 & 0.785 & 1.415 & 1.347 & 2.088 & 0.801 & 1.258 & 1.357 & 1.937 \\
    JEANet & 0.979 & \textbf{2.170} & 1.426 & $\textbf{2.883}^{\dagger}$ & 0.792 & \textbf{1.464} & \textbf{1.362} & $\textbf{2.145}^{\dagger}$ & 0.936 & \textbf{1.445} & \textbf{1.486} & $\textbf{2.188}^{\dagger}$ \\
    \bottomrule
  \end{tabular}
\end{table*}

\begin{table*}
\centering
  \caption{Experimental results for the different value distributions. Each VCG metric is set to 1, with $score$ calculated as 1.5, and each column is normalized. The regret of mechanisms generated by Revised JRegNet, Revised TICNet, and JEANet are < 0.001. The best performance is highlighted in bold. ``$\dagger$'' indicates a statistically significant improvement in a paired $t$-test at $p$ < 0.05.}
  \label{tab:third table}
  \begin{tabular}{ccccccccccccc}
    \toprule
    \multirow{2}{*}{Method} & \multicolumn{4}{c}{Uniform} & \multicolumn{4}{c}{Normal} & \multicolumn{4}{c}{Lognormal} \\
    & SW & Rev & UE & Score & SW & Rev & UE & Score & SW & Rev & UE & Score \\
    \midrule
    GSP with Fixed Positions & 0.789 & 1.592 & 1.248 & 2.216 & 0.506 & 1.478 & 1.399 & 2.178 & 0.787 & 1.326 & 1.253 & 1.953 \\
    VCG & \textbf{1.000} & 1.000 & 1.000 & 1.500 & \textbf{1.000} & 1.000 & 1.000 & 1.500 & \textbf{1.000} & 1.000 & 1.000 & 1.500 \\
    IAS & 0.375 & 0.745 & 1.150 & 1.320 & 0.438 & 1.427 & 1.137 & 1.996 & 0.822 & 1.664 & 1.064 & 2.196 \\
    Revised JRegNet & 0.716 & 1.444 & 1.191 & 2.040 & 0.446 & 1.537 & 1.339 & 2.206 & 0.768 & 1.473 & 1.189 & 2.067 \\
    Revised TICNet & 0.636 & 1.845 & 1.484 & 2.587 & 0.339 & 1.511 & 1.415 & 2.219 & 0.800 & \textbf{1.848} & 1.317 & 2.507 \\
    JEANet & 0.722 & \textbf{2.059} & \textbf{1.520} & $\textbf{2.820}^{\dagger}$ & 0.447 & \textbf{1.669} & \textbf{1.476} & $\textbf{2.407}^{\dagger}$ & 0.834 & 1.833 & \textbf{1.414} & $\textbf{2.540}^{\dagger}$ \\
    \bottomrule
  \end{tabular}
\end{table*}

\begin{itemize}[leftmargin=*]
\item \textbf{Industrial Dataset}. Beyond the synthetic dataset, we experimented with real online request logs. The industrial dataset (March 1–15, 2025) comes from the platform’s GSP auctions and fixed positions. Each sample includes candidate ads, organic items, bid profiles, and user experience profiles. We divided the March 1–14 data into a training set, and the March 15 data serves as a test set. Due to the varying numbers of advertisers, organic items, and slots in the logs, we pruned the data for valid setup.
\end{itemize}

\subsubsection{Evaluation Metrics} To assess the performance of JEANet and baselines, we evaluate the following empirical metrics: social welfare: $\text{SW} = \frac{1}{L} \sum_{\ell = 1}^{L} \sum_{i = 1}^{m} v_i^{(\ell)} \cdot a_i^w$, revenue: $\text{Rev} = \frac{1}{L} \sum_{\ell = 1}^{L} \sum_{i = 1}^{m} p_i^w$, user experience: $\text{UE} = \frac{1}{L} \sum_{\ell = 1}^{L} \sum_{i = 1}^{m + n} ue_i \cdot a_i^w$, and the combination of revenue and user experience with coefficient $\gamma$: $Score = \text{Rev} + \gamma \text{UE} = \frac{1}{L} \sum_{\ell = 1}^{L} \left[ \sum_{i = 1}^{m} p_i^w + \gamma \sum_{i = 1}^{m + n} ue_i \cdot a_i^w\right]$. For the $Score$ metric, the hyperparameter $\gamma$, validated through a series of experiments with the range [0,1], performs best when set to 0.5.

\begin{table*}[h]
\centering
  \caption{Experimental results for the industrial data. Each VCG metric is set to 1, with $score$ calculated as 1.5, and each column is normalized. The regret of mechanisms generated by Revised JRegNet, Revised TICNet, and JEANet are < 0.001. The best performance is highlighted in bold. ``$\dagger$'' indicates a statistically significant improvement in a paired $t$-test at $p$ < 0.05 level.}
  \label{tab:fourth table}
  \begin{tabular}{ccccccccccccc}
    \toprule
    \multirow{2}{*}{Method} & \multicolumn{4}{c}{slot=4} & \multicolumn{4}{c}{slot=5} & \multicolumn{4}{c}{slot=6} \\
    & SW & Rev & UE & Score & SW & Rev & UE &Score & SW & Rev & UE & Score \\
    \midrule
    GSP with Fixed Positions & 0.818 & 1.025 & 2.640 & 2.345 & 0.862 & 1.142 & 2.347 & 2.316 & 0.890 & 1.264 & 2.124 & 2.326 \\
    VCG & \textbf{1.000} & 1.000 & 1.000 & 1.500 & \textbf{1.000} & 1.000 & 1.000 & 1.500 & \textbf{1.000} & 1.000 & 1.000 & 1.500 \\
    IAS & 0.998 & 1.208 & 3.080 & 2.748 & 0.998 & 1.193 & 2.724 & 2.556 & 0.998 & 1.199 & 2.412 & 2.405 \\
    Revised JRegNet & 0.717 & 0.830 & 3.000 & 2.330 & 0.884 & 1.230 & 2.745 & 2.603 & 0.913 & 1.348 & 2.454 & 2.575 \\
    Revised TICNet & 0.965 & 1.307 & 3.610 & 3.112 & 0.992 & 1.300 & \textbf{3.704} & 3.152 & 0.903 & 1.269 & \textbf{3.608} & 3.073 \\
    JEANet & 0.953 & \textbf{1.350} & \textbf{3.690} & $\textbf{3.195}^{\dagger}$ & 0.958 & \textbf{1.375} & 3.653 & $\textbf{3.201}^{\dagger}$ & 0.882 & \textbf{1.357} & 3.598 & $\textbf{3.156}^{\dagger}$ \\
    \bottomrule
  \end{tabular}
\end{table*}

\subsubsection{Baseline Methods} We compare JEANet with the following representative mechanisms for integrating ad auction and recommendation system:

\begin{itemize}[leftmargin=*]
\item \textbf{GSP \cite{edelman2007internet} with Fixed Positions}. This well-known method proceeds sequentially: First, determine the number of ad slots and their positions. Next, rank organic items by user experience. Then set ads order via the GSP mechanism. Finally, place winning ads in preset slots and charge accordingly.
\end{itemize}

\begin{itemize}[leftmargin=*]
\item \textbf{VCG \cite{clarke1971multipart,groves1973incentives,vickrey1961counterspeculation}}. It is a classic mechanism that satisfies DSIC and IR while maximizing social welfare.
\end{itemize}

\begin{itemize}[leftmargin=*]
\item \textbf{IAS \cite{li2023optimally}}. A Myerson-based mechanism ranks ads and organic items using a ranking score of $\phi_i(v_i) + \gamma \cdot ue_i$ , where $\phi_i(v_i) = v_i - \frac{1 - F_i(v_i)}{f_i(v_i)}$ is the virtual value. Payments are determined according to the Myerson payment rule.
\end{itemize}

\begin{itemize}[leftmargin=*]
\item \textbf{Revised JRegNet \cite{zhang2024joint}}. Based on JRegNet, revised JRegNet aims to output a hybrid allocation of stores, brands, and joints. In detail, a dummy brand with value 0 is created to link to all stores, and generated bundles represent independent.
\end{itemize}

\begin{itemize}[leftmargin=*]
\item \textbf{TICNet \cite{ma2025context}}. This is an auction mechanism that balances revenue and user experience by considering contextual information.
\end{itemize}

Note that TICNet only allows brands or stores in auctions. We modified it to let joint ads also enter, with consolidated payments.

\subsection{Experimental results}
\subsubsection{Synthetic Dataset Experiment (\textbf{RQ1})}
To further validate JEANet's superiority, we conduct experiments across various settings, including different slot numbers and ad-organic items ratio.

\begin{itemize}[leftmargin=*]
\item \textbf{Random number of items}. This random-item experimental design simulates real online environments and ensures fair comparison of methods. The detailed results are shown in Table \ref{tab:first table}. A collective analysis of baselines shows that JEANet achieves the highest $Score$ in all experiments and nearly highest $UE$, despite social welfare loss compared with VCG. This indicates JEANet's ETM effectively captures externalities, balancing $Rev$ and $UE$ better than baselines with identical ad load and regret <0.001.
\end{itemize}

\begin{itemize}[leftmargin=*]
\item \textbf{Fixed number of items}. The results of thie fixed-item experiment for settings A, B, and C are shown in Table ~\ref{tab:second table}. In settings with varying ads to organic items proportions, the mechanisms computed by JEANet achieves significantly higher $Rev$ and $Score$ than others, with regret <0.001. This confirms JEANet's effectiveness in generating globally optimal hybrid lists.
\end{itemize}

\begin{itemize}[leftmargin=*]
\item \textbf{Different Value Distributions}. To demonstrate JEANet’s generalization ability, we conducted experiments under diverse bid distributions in setting A: uniform $U[0, 1]$, truncated normal $N(0.5, 0.5)$ with $v \in [0, 1]$, and truncated multi-lognormal ($LN_1 \break (0.3,0.2)$, $LN_2(0.5,0.3)$, $LN_3(0.7,0.4)$, $LN_4(0.9,0.5)$) with $v \in [0, 1]$. Table ~\ref{tab:third table} shows that JEANet consistently achieves the highest $Score$ across distributions, outperforming other mechanisms. This highlights its robustness with diverse bids and validates the effectiveness of JEANet’s AEM module for feature extraction.
\end{itemize}


\subsubsection{Industrial Dataset Experiment (\textbf{RQ2})} Industrial dataset results are in Table ~\ref{tab:fourth table}. Compared with baselines, the mechanisms computed by JEANet achieve significantly higher $Rev$ and $Score$, and are closest to the highest $UE$, supported by paired $t$-tests ($p < 0.05$). Additionally, the mechanisms nearly satisfy DSIC (regret $<0.001$).

Furthermore, JEANet is deployable in real ad scenarios. On a 64-CPUs server, 30,000 offline training iterations take 3–4 hours. The trained JEANet’s forward propagation is extremely fast ($< 10$ milliseconds per sample), leaving online decision-making unaffected. This underscores its effectiveness in industrial auctions and feasibility for deployment in real ad scenarios.

\subsubsection{IC Discussion in Real Industrial Scenarios (\textbf{RQ3})} Although the mechanism computed by JEANet is an approximate DSIC mechanism lacking an exact strategy-proofness guarantee, in practical scenarios, we consider two points:
\begin{itemize}[leftmargin=*]
\item \textbf{Real scenarios differ from theoretical ones.} Although theory assumes advertisers bid to maximize utility, in real ad scenarios, they rarely find the optimal misreporting strategy due to costs, using suboptimal bids instead. Thus, approximate-IC nearly guarantees truthful bidding in practice, given misreporting costs.
\end{itemize}

\begin{itemize}[leftmargin=*]
\item \textbf{Relaxing the DSIC constraints can yield higher revenue.} In real advertising scenarios, approximate-IC may achieve higher revenue than exact-IC by sacrificing a certain degree of IC.
\end{itemize}

\subsubsection{Ablation Study (\textbf{RQ4})}
To validate the effectiveness of different components, we conducted experiments under Setting (A) using the synthetic dataset. Correspondingly, we designed ablation studies with two variants that simplify JEANet in different ways:
\begin{itemize}[leftmargin=*]
\item \textbf{ETM+DMM} removes the Adaptive Extraction Module (AEM), using original bid profiles directly as features. This setting verifies AEM’s effectiveness.
\end{itemize}

\begin{itemize}[leftmargin=*]
\item \textbf{MLP+DMM} removes the Adaptive Extraction Module (AEM) and replaces the Externality Transformer Module (ETM) with a 3-layer MLP, with layer parameters [64, 32, 16]. This comparison assesses ETM’s benefit in capturing global externalities.
\end{itemize}

\begin{table}
  \caption{Ablation Study of JEANet. Each JEANet metric is set to 1, with $score$ calculated as 1.5, and each column is normalized. The regret of all mechanisms are < 0.001. The best performance is highlighted in bold. ``$\dagger$'' indicates a statistically significant improvement in a paired $t$-test at $p$ < 0.05.}
  \label{tab:fifth table}
  \begin{tabular}{cccc}
    \toprule
    Method&Rev&UE&Score\\
    \midrule
    AEM+ETM+DMM & \textbf{1.000} & \textbf{1.000} & \textbf{1.500}\\
    ETM+DMM & 0.942(-5.8\%) & 0.988(-1.2\%) & 1.435(-4.3\%)\\
    MLP+DMM & 0.911(-8.9\%) & 0.958(-4.2\%) & 1.390(-7.4\%)\\
  \bottomrule
\end{tabular}
\end{table}

Performance results in Table ~\ref{tab:fifth table} show that the full JEANet (comprising \textbf{AEM}, \textbf{ETM}, and \textbf{DMM}) consistently outperforms all ablation variants across major metrics, validating the effectiveness of our design. Specifically,

\begin{enumerate}[label=(\arabic*),leftmargin=*]
    \item \textbf{ETM+DMM} significantly reduces $Rev$ (-5.8\%), $UE$ (-1.2\%), and $Score$ (-4.3\%), highlighting AEM’s importance. In particular, these overall improvements indicate that AEM effectively reflects the bid distribution characteristics of different advertisers.
    \item \textbf{MLP+DMM} variant degrades performance, with $Rev$ (-8.9\%), $UE$ (-4.2\%) and $Score$ (-7.4\%). This highlights ETM’s effectiveness in balancing performance and externality modeling efficiency. Moreover, the transformer-based ETM outperforms the MLP in maintaining mechanism anonymity.
\end{enumerate}

In summary, JEANet exhibits the same trends seen in other datasets and settings. Each component of JEANet, including the AEM and the ETM, plays an important role in improving revenue and user experience, while ensuring economic robustness.

\section{Conclusion}
In this article, we focus on designing optimal joint advertising mechanisms with externalities and adaptation that seamlessly integrate ads and organic items into hybrid listings, including store ads, brand ads, and joint ads. To address this challenge, we propose JEANet, a neural network architecture based on RQ and Transformer. JEANet features an AEM to extract multi-distribution bid features from advertisers, an ETM to model global cross-item externalities, and compatibility with both traditional and joint advertising frameworks. Through comprehensive experiments, we demonstrate JEANet’s superiority over baselines. 

\bibliographystyle{ACM-Reference-Format}
\bibliography{sample-base}


\begin{thebibliography}{43}


\ifx \showCODEN    \undefined \def \showCODEN     #1{\unskip}     \fi
\ifx \showISBNx    \undefined \def \showISBNx     #1{\unskip}     \fi
\ifx \showISBNxiii \undefined \def \showISBNxiii  #1{\unskip}     \fi
\ifx \showISSN     \undefined \def \showISSN      #1{\unskip}     \fi
\ifx \showLCCN     \undefined \def \showLCCN      #1{\unskip}     \fi
\ifx \shownote     \undefined \def \shownote      #1{#1}          \fi
\ifx \showarticletitle \undefined \def \showarticletitle #1{#1}   \fi
\ifx \showURL      \undefined \def \showURL       {\relax}        \fi
\providecommand\bibfield[2]{#2}
\providecommand\bibinfo[2]{#2}
\providecommand\natexlab[1]{#1}
\providecommand\showeprint[2][]{arXiv:#2}

\bibitem[Aggarwal et~al\mbox{.}(2024)]%
        {aggarwal2024selling}
\bibfield{author}{\bibinfo{person}{Gagan Aggarwal}, \bibinfo{person}{Ashwinkumar Badanidiyuru}, \bibinfo{person}{Paul D{\"u}tting}, {and} \bibinfo{person}{Federico Fusco}.} \bibinfo{year}{2024}\natexlab{}.
\newblock \showarticletitle{Selling joint ads: A regret minimization perspective}. In \bibinfo{booktitle}{\emph{Proceedings of the 25th ACM Conference on Economics and Computation}}. \bibinfo{pages}{164--194}.
\newblock


\bibitem[Balcan et~al\mbox{.}(2008)]%
        {balcan2008reducing}
\bibfield{author}{\bibinfo{person}{Maria-Florina Balcan}, \bibinfo{person}{Avrim Blum}, \bibinfo{person}{Jason~D Hartline}, {and} \bibinfo{person}{Yishay Mansour}.} \bibinfo{year}{2008}\natexlab{}.
\newblock \showarticletitle{Reducing mechanism design to algorithm design via machine learning}.
\newblock \bibinfo{journal}{\emph{J. Comput. System Sci.}} \bibinfo{volume}{74}, \bibinfo{number}{8} (\bibinfo{year}{2008}), \bibinfo{pages}{1245--1270}.
\newblock


\bibitem[Birkhoff(1946)]%
        {birkhoff1946tres}
\bibfield{author}{\bibinfo{person}{Garrett Birkhoff}.} \bibinfo{year}{1946}\natexlab{}.
\newblock \showarticletitle{Tres observaciones sobre el algebra lineal}.
\newblock \bibinfo{journal}{\emph{Univ. Nac. Tucuman, Ser. A}}  \bibinfo{volume}{5} (\bibinfo{year}{1946}), \bibinfo{pages}{147--154}.
\newblock


\bibitem[Caragiannis et~al\mbox{.}(2011)]%
        {caragiannis2011efficiency}
\bibfield{author}{\bibinfo{person}{Ioannis Caragiannis}, \bibinfo{person}{Christos Kaklamanis}, \bibinfo{person}{Panagiotis Kanellopoulos}, {and} \bibinfo{person}{Maria Kyropoulou}.} \bibinfo{year}{2011}\natexlab{}.
\newblock \showarticletitle{On the efficiency of equilibria in generalized second price auctions}. In \bibinfo{booktitle}{\emph{Proceedings of the 12th ACM conference on Electronic commerce}}. \bibinfo{pages}{81--90}.
\newblock


\bibitem[Clarke(1971)]%
        {clarke1971multipart}
\bibfield{author}{\bibinfo{person}{Edward~H Clarke}.} \bibinfo{year}{1971}\natexlab{}.
\newblock \showarticletitle{Multipart pricing of public goods}.
\newblock \bibinfo{journal}{\emph{Public choice}} (\bibinfo{year}{1971}), \bibinfo{pages}{17--33}.
\newblock


\bibitem[Conitzer and Sandholm(2002)]%
        {conitzer2002complexity}
\bibfield{author}{\bibinfo{person}{Vincent Conitzer} {and} \bibinfo{person}{Tuomas Sandholm}.} \bibinfo{year}{2002}\natexlab{}.
\newblock \showarticletitle{Complexity of mechanism design}.
\newblock \bibinfo{journal}{\emph{arXiv preprint cs/0205075}} (\bibinfo{year}{2002}).
\newblock


\bibitem[Conitzer and Sandholm(2004)]%
        {conitzer2004self}
\bibfield{author}{\bibinfo{person}{Vincent Conitzer} {and} \bibinfo{person}{Tuomas Sandholm}.} \bibinfo{year}{2004}\natexlab{}.
\newblock \showarticletitle{Self-interested automated mechanism design and implications for optimal combinatorial auctions}. In \bibinfo{booktitle}{\emph{Proceedings of the 5th ACM Conference on Electronic Commerce}}. \bibinfo{pages}{132--141}.
\newblock


\bibitem[Duan et~al\mbox{.}(2022)]%
        {duan2022context}
\bibfield{author}{\bibinfo{person}{Zhijian Duan}, \bibinfo{person}{Jingwu Tang}, \bibinfo{person}{Yutong Yin}, \bibinfo{person}{Zhe Feng}, \bibinfo{person}{Xiang Yan}, \bibinfo{person}{Manzil Zaheer}, {and} \bibinfo{person}{Xiaotie Deng}.} \bibinfo{year}{2022}\natexlab{}.
\newblock \showarticletitle{A context-integrated transformer-based neural network for auction design}. In \bibinfo{booktitle}{\emph{International Conference on Machine Learning}}. PMLR, \bibinfo{pages}{5609--5626}.
\newblock


\bibitem[D{\"u}tting et~al\mbox{.}(2024)]%
        {dutting2024optimal}
\bibfield{author}{\bibinfo{person}{Paul D{\"u}tting}, \bibinfo{person}{Zhe Feng}, \bibinfo{person}{Harikrishna Narasimhan}, \bibinfo{person}{David~C Parkes}, {and} \bibinfo{person}{Sai~Srivatsa Ravindranath}.} \bibinfo{year}{2024}\natexlab{}.
\newblock \showarticletitle{Optimal auctions through deep learning: Advances in differentiable economics}.
\newblock \bibinfo{journal}{\emph{J. ACM}} \bibinfo{volume}{71}, \bibinfo{number}{1} (\bibinfo{year}{2024}), \bibinfo{pages}{1--53}.
\newblock


\bibitem[D{\"u}tting et~al\mbox{.}(2015)]%
        {dutting2015payment}
\bibfield{author}{\bibinfo{person}{Paul D{\"u}tting}, \bibinfo{person}{Felix Fischer}, \bibinfo{person}{Pichayut Jirapinyo}, \bibinfo{person}{John~K Lai}, \bibinfo{person}{Benjamin Lubin}, {and} \bibinfo{person}{David~C Parkes}.} \bibinfo{year}{2015}\natexlab{}.
\newblock \bibinfo{title}{Payment rules through discriminant-based classifiers}.
\newblock


\bibitem[Edelman et~al\mbox{.}(2007)]%
        {edelman2007internet}
\bibfield{author}{\bibinfo{person}{Benjamin Edelman}, \bibinfo{person}{Michael Ostrovsky}, {and} \bibinfo{person}{Michael Schwarz}.} \bibinfo{year}{2007}\natexlab{}.
\newblock \showarticletitle{Internet advertising and the generalized second-price auction: Selling billions of dollars worth of keywords}.
\newblock \bibinfo{journal}{\emph{American economic review}} \bibinfo{volume}{97}, \bibinfo{number}{1} (\bibinfo{year}{2007}), \bibinfo{pages}{242--259}.
\newblock


\bibitem[Feng et~al\mbox{.}(2018)]%
        {feng2018deep}
\bibfield{author}{\bibinfo{person}{Zhe Feng}, \bibinfo{person}{Harikrishna Narasimhan}, {and} \bibinfo{person}{David~C Parkes}.} \bibinfo{year}{2018}\natexlab{}.
\newblock \showarticletitle{Deep learning for revenue-optimal auctions with budgets}. In \bibinfo{booktitle}{\emph{Proceedings of the 17th international conference on autonomous agents and multiagent systems}}. \bibinfo{pages}{354--362}.
\newblock


\bibitem[Gatti et~al\mbox{.}(2012)]%
        {gatti2012truthful}
\bibfield{author}{\bibinfo{person}{Nicola Gatti}, \bibinfo{person}{Alessandro Lazaric}, {and} \bibinfo{person}{Francesco Trovo}.} \bibinfo{year}{2012}\natexlab{}.
\newblock \showarticletitle{A truthful learning mechanism for contextual multi-slot sponsored search auctions with externalities}. In \bibinfo{booktitle}{\emph{Proceedings of the 13th ACM Conference on Electronic Commerce}}. \bibinfo{pages}{605--622}.
\newblock


\bibitem[Ghosh and Mahdian(2008)]%
        {ghosh2008externalities}
\bibfield{author}{\bibinfo{person}{Arpita Ghosh} {and} \bibinfo{person}{Mohammad Mahdian}.} \bibinfo{year}{2008}\natexlab{}.
\newblock \showarticletitle{Externalities in online advertising}. In \bibinfo{booktitle}{\emph{Proceedings of the 17th international conference on World Wide Web}}. \bibinfo{pages}{161--168}.
\newblock


\bibitem[Gomes and Sweeney(2009)]%
        {gomes2009bayes}
\bibfield{author}{\bibinfo{person}{Renato~D Gomes} {and} \bibinfo{person}{Kane~S Sweeney}.} \bibinfo{year}{2009}\natexlab{}.
\newblock \showarticletitle{Bayes-Nash equilibria of the generalized second price auction}. In \bibinfo{booktitle}{\emph{Proceedings of the 10th ACM conference on Electronic Commerce}}. \bibinfo{pages}{107--108}.
\newblock


\bibitem[Groves(1973)]%
        {groves1973incentives}
\bibfield{author}{\bibinfo{person}{Theodore Groves}.} \bibinfo{year}{1973}\natexlab{}.
\newblock \showarticletitle{Incentives in teams}.
\newblock \bibinfo{journal}{\emph{Econometrica: Journal of the Econometric Society}} (\bibinfo{year}{1973}), \bibinfo{pages}{617--631}.
\newblock


\bibitem[Huang et~al\mbox{.}(2021)]%
        {huang2021deep}
\bibfield{author}{\bibinfo{person}{Jianqiang Huang}, \bibinfo{person}{Ke Hu}, \bibinfo{person}{Qingtao Tang}, \bibinfo{person}{Mingjian Chen}, \bibinfo{person}{Yi Qi}, \bibinfo{person}{Jia Cheng}, {and} \bibinfo{person}{Jun Lei}.} \bibinfo{year}{2021}\natexlab{}.
\newblock \showarticletitle{Deep position-wise interaction network for ctr prediction}. In \bibinfo{booktitle}{\emph{Proceedings of the 44th International ACM SIGIR Conference on Research and Development in Information Retrieval}}. \bibinfo{pages}{1885--1889}.
\newblock


\bibitem[Ivanov et~al\mbox{.}(2022)]%
        {ivanov2022optimal}
\bibfield{author}{\bibinfo{person}{Dmitry Ivanov}, \bibinfo{person}{Iskander Safiulin}, \bibinfo{person}{Igor Filippov}, {and} \bibinfo{person}{Ksenia Balabaeva}.} \bibinfo{year}{2022}\natexlab{}.
\newblock \showarticletitle{Optimal-er auctions through attention}.
\newblock \bibinfo{journal}{\emph{Advances in Neural Information Processing Systems}}  \bibinfo{volume}{35} (\bibinfo{year}{2022}), \bibinfo{pages}{34734--34747}.
\newblock


\bibitem[Kuo et~al\mbox{.}(2020)]%
        {kuo2020proportionnet}
\bibfield{author}{\bibinfo{person}{Kevin Kuo}, \bibinfo{person}{Anthony Ostuni}, \bibinfo{person}{Elizabeth Horishny}, \bibinfo{person}{Michael~J Curry}, \bibinfo{person}{Samuel Dooley}, \bibinfo{person}{Ping-yeh Chiang}, \bibinfo{person}{Tom Goldstein}, {and} \bibinfo{person}{John~P Dickerson}.} \bibinfo{year}{2020}\natexlab{}.
\newblock \showarticletitle{Proportionnet: Balancing fairness and revenue for auction design with deep learning}.
\newblock \bibinfo{journal}{\emph{arXiv preprint arXiv:2010.06398}} (\bibinfo{year}{2020}).
\newblock


\bibitem[Lahaie(2011)]%
        {lahaie2011kernel}
\bibfield{author}{\bibinfo{person}{S{\'e}bastien Lahaie}.} \bibinfo{year}{2011}\natexlab{}.
\newblock \showarticletitle{A kernel-based iterative combinatorial auction}. In \bibinfo{booktitle}{\emph{Proceedings of the AAAI Conference on Artificial Intelligence}}, Vol.~\bibinfo{volume}{25}. \bibinfo{pages}{695--700}.
\newblock


\bibitem[Lahaie and Pennock(2007)]%
        {lahaie2007revenue}
\bibfield{author}{\bibinfo{person}{S{\'e}bastien Lahaie} {and} \bibinfo{person}{David~M Pennock}.} \bibinfo{year}{2007}\natexlab{}.
\newblock \showarticletitle{Revenue analysis of a family of ranking rules for keyword auctions}. In \bibinfo{booktitle}{\emph{Proceedings of the 8th ACM Conference on Electronic Commerce}}. \bibinfo{pages}{50--56}.
\newblock


\bibitem[Lee et~al\mbox{.}(2022)]%
        {lee2022autoregressive}
\bibfield{author}{\bibinfo{person}{Doyup Lee}, \bibinfo{person}{Chiheon Kim}, \bibinfo{person}{Saehoon Kim}, \bibinfo{person}{Minsu Cho}, {and} \bibinfo{person}{Wook-Shin Han}.} \bibinfo{year}{2022}\natexlab{}.
\newblock \showarticletitle{Autoregressive image generation using residual quantization}. In \bibinfo{booktitle}{\emph{Proceedings of the IEEE/CVF Conference on Computer Vision and Pattern Recognition}}. \bibinfo{pages}{11523--11532}.
\newblock


\bibitem[Li et~al\mbox{.}(2023)]%
        {li2023optimally}
\bibfield{author}{\bibinfo{person}{Weian Li}, \bibinfo{person}{Qi Qi}, \bibinfo{person}{Changjun Wang}, {and} \bibinfo{person}{Changyuan Yu}.} \bibinfo{year}{2023}\natexlab{}.
\newblock \showarticletitle{Optimally integrating ad auction into E-commerce platforms}.
\newblock \bibinfo{journal}{\emph{Theoretical Computer Science}}  \bibinfo{volume}{976} (\bibinfo{year}{2023}), \bibinfo{pages}{114141}.
\newblock


\bibitem[Li et~al\mbox{.}(2024)]%
        {li2024deep}
\bibfield{author}{\bibinfo{person}{Xuejian Li}, \bibinfo{person}{Ze Wang}, \bibinfo{person}{Bingqi Zhu}, \bibinfo{person}{Fei He}, \bibinfo{person}{Yongkang Wang}, {and} \bibinfo{person}{Xingxing Wang}.} \bibinfo{year}{2024}\natexlab{}.
\newblock \showarticletitle{Deep automated mechanism design for integrating ad auction and allocation in feed}. In \bibinfo{booktitle}{\emph{Proceedings of the 47th International ACM SIGIR Conference on Research and Development in Information Retrieval}}. \bibinfo{pages}{1211--1220}.
\newblock


\bibitem[Liao et~al\mbox{.}(2022a)]%
        {liao2022nma}
\bibfield{author}{\bibinfo{person}{Guogang Liao}, \bibinfo{person}{Xuejian Li}, \bibinfo{person}{Ze Wang}, \bibinfo{person}{Fan Yang}, \bibinfo{person}{Muzhi Guan}, \bibinfo{person}{Bingqi Zhu}, \bibinfo{person}{Yongkang Wang}, \bibinfo{person}{Xingxing Wang}, {and} \bibinfo{person}{Dong Wang}.} \bibinfo{year}{2022}\natexlab{a}.
\newblock \showarticletitle{NMA: neural multi-slot auctions with externalities for online advertising}.
\newblock \bibinfo{journal}{\emph{arXiv preprint arXiv:2205.10018}} (\bibinfo{year}{2022}).
\newblock


\bibitem[Liao et~al\mbox{.}(2022b)]%
        {liao2022cross}
\bibfield{author}{\bibinfo{person}{Guogang Liao}, \bibinfo{person}{Ze Wang}, \bibinfo{person}{Xiaoxu Wu}, \bibinfo{person}{Xiaowen Shi}, \bibinfo{person}{Chuheng Zhang}, \bibinfo{person}{Yongkang Wang}, \bibinfo{person}{Xingxing Wang}, {and} \bibinfo{person}{Dong Wang}.} \bibinfo{year}{2022}\natexlab{b}.
\newblock \showarticletitle{Cross dqn: Cross deep q network for ads allocation in feed}. In \bibinfo{booktitle}{\emph{Proceedings of the ACM Web Conference 2022}}. \bibinfo{pages}{401--409}.
\newblock


\bibitem[Liu et~al\mbox{.}(2021)]%
        {liu2021neural}
\bibfield{author}{\bibinfo{person}{Xiangyu Liu}, \bibinfo{person}{Chuan Yu}, \bibinfo{person}{Zhilin Zhang}, \bibinfo{person}{Zhenzhe Zheng}, \bibinfo{person}{Yu Rong}, \bibinfo{person}{Hongtao Lv}, \bibinfo{person}{Da Huo}, \bibinfo{person}{Yiqing Wang}, \bibinfo{person}{Dagui Chen}, \bibinfo{person}{Jian Xu}, {et~al\mbox{.}}} \bibinfo{year}{2021}\natexlab{}.
\newblock \showarticletitle{Neural auction: End-to-end learning of auction mechanisms for e-commerce advertising}. In \bibinfo{booktitle}{\emph{Proceedings of the 27th ACM SIGKDD Conference on Knowledge Discovery \& Data Mining}}. \bibinfo{pages}{3354--3364}.
\newblock


\bibitem[Lucier et~al\mbox{.}(2012)]%
        {lucier2012revenue}
\bibfield{author}{\bibinfo{person}{Brendan Lucier}, \bibinfo{person}{Renato Paes~Leme}, {and} \bibinfo{person}{{\'E}va Tardos}.} \bibinfo{year}{2012}\natexlab{}.
\newblock \showarticletitle{On revenue in the generalized second price auction}. In \bibinfo{booktitle}{\emph{Proceedings of the 21st international conference on World Wide Web}}. \bibinfo{pages}{361--370}.
\newblock


\bibitem[Ma et~al\mbox{.}(2025)]%
        {ma2025context}
\bibfield{author}{\bibinfo{person}{Yuchao Ma}, \bibinfo{person}{Weian Li}, \bibinfo{person}{Yuejia Dou}, \bibinfo{person}{Zhiyuan Su}, \bibinfo{person}{Changyuan Yu}, {and} \bibinfo{person}{Qi Qi}.} \bibinfo{year}{2025}\natexlab{}.
\newblock \showarticletitle{A Context-Aware Framework for Integrating Ad Auctions and Recommendations}. In \bibinfo{booktitle}{\emph{Proceedings of the ACM on Web Conference 2025}}. \bibinfo{pages}{1977--1991}.
\newblock


\bibitem[Ma et~al\mbox{.}(2024)]%
        {ma2024joint}
\bibfield{author}{\bibinfo{person}{Yuchao Ma}, \bibinfo{person}{Weian Li}, \bibinfo{person}{Wanzhi Zhang}, \bibinfo{person}{Yahui Lei}, \bibinfo{person}{Zhicheng Zhang}, \bibinfo{person}{Qi Qi}, \bibinfo{person}{Qiang Liu}, {and} \bibinfo{person}{Xingxing Wang}.} \bibinfo{year}{2024}\natexlab{}.
\newblock \showarticletitle{Joint Bidding in Ad Auctions}. In \bibinfo{booktitle}{\emph{Annual Conference on Theory and Applications of Models of Computation}}. Springer, \bibinfo{pages}{344--354}.
\newblock


\bibitem[Myerson(1981)]%
        {myerson1981optimal}
\bibfield{author}{\bibinfo{person}{Roger~B Myerson}.} \bibinfo{year}{1981}\natexlab{}.
\newblock \showarticletitle{Optimal auction design}.
\newblock \bibinfo{journal}{\emph{Mathematics of operations research}} \bibinfo{volume}{6}, \bibinfo{number}{1} (\bibinfo{year}{1981}), \bibinfo{pages}{58--73}.
\newblock


\bibitem[Peri et~al\mbox{.}(2021)]%
        {peri2021preferencenet}
\bibfield{author}{\bibinfo{person}{Neehar Peri}, \bibinfo{person}{Michael Curry}, \bibinfo{person}{Samuel Dooley}, {and} \bibinfo{person}{John Dickerson}.} \bibinfo{year}{2021}\natexlab{}.
\newblock \showarticletitle{Preferencenet: Encoding human preferences in auction design with deep learning}.
\newblock \bibinfo{journal}{\emph{Advances in Neural Information Processing Systems}}  \bibinfo{volume}{34} (\bibinfo{year}{2021}), \bibinfo{pages}{17532--17542}.
\newblock


\bibitem[Qin et~al\mbox{.}(2015)]%
        {qin2015sponsored}
\bibfield{author}{\bibinfo{person}{Tao Qin}, \bibinfo{person}{Wei Chen}, {and} \bibinfo{person}{Tie-Yan Liu}.} \bibinfo{year}{2015}\natexlab{}.
\newblock \showarticletitle{Sponsored search auctions: Recent advances and future directions}.
\newblock \bibinfo{journal}{\emph{ACM Transactions on Intelligent Systems and Technology (TIST)}} \bibinfo{volume}{5}, \bibinfo{number}{4} (\bibinfo{year}{2015}), \bibinfo{pages}{1--34}.
\newblock


\bibitem[Rahme et~al\mbox{.}(2021)]%
        {rahme2021permutation}
\bibfield{author}{\bibinfo{person}{Jad Rahme}, \bibinfo{person}{Samy Jelassi}, \bibinfo{person}{Joan Bruna}, {and} \bibinfo{person}{S~Matthew Weinberg}.} \bibinfo{year}{2021}\natexlab{}.
\newblock \showarticletitle{A permutation-equivariant neural network architecture for auction design}. In \bibinfo{booktitle}{\emph{Proceedings of the AAAI conference on artificial intelligence}}, Vol.~\bibinfo{volume}{35}. \bibinfo{pages}{5664--5672}.
\newblock


\bibitem[Sandholm and Likhodedov(2015)]%
        {sandholm2015automated}
\bibfield{author}{\bibinfo{person}{Tuomas Sandholm} {and} \bibinfo{person}{Anton Likhodedov}.} \bibinfo{year}{2015}\natexlab{}.
\newblock \showarticletitle{Automated design of revenue-maximizing combinatorial auctions}.
\newblock \bibinfo{journal}{\emph{Operations Research}} \bibinfo{volume}{63}, \bibinfo{number}{5} (\bibinfo{year}{2015}), \bibinfo{pages}{1000--1025}.
\newblock


\bibitem[Thompson and Leyton-Brown(2013)]%
        {thompson2013revenue}
\bibfield{author}{\bibinfo{person}{David~RM Thompson} {and} \bibinfo{person}{Kevin Leyton-Brown}.} \bibinfo{year}{2013}\natexlab{}.
\newblock \showarticletitle{Revenue optimization in the generalized second-price auction}. In \bibinfo{booktitle}{\emph{Proceedings of the fourteenth ACM conference on Electronic commerce}}. \bibinfo{pages}{837--852}.
\newblock


\bibitem[Vickrey(1961)]%
        {vickrey1961counterspeculation}
\bibfield{author}{\bibinfo{person}{William Vickrey}.} \bibinfo{year}{1961}\natexlab{}.
\newblock \showarticletitle{Counterspeculation, auctions, and competitive sealed tenders}.
\newblock \bibinfo{journal}{\emph{The Journal of finance}} \bibinfo{volume}{16}, \bibinfo{number}{1} (\bibinfo{year}{1961}), \bibinfo{pages}{8--37}.
\newblock


\bibitem[Wang et~al\mbox{.}(2024)]%
        {wang2024gemnet}
\bibfield{author}{\bibinfo{person}{Tonghan Wang}, \bibinfo{person}{Yanchen Jiang}, {and} \bibinfo{person}{David~C Parkes}.} \bibinfo{year}{2024}\natexlab{}.
\newblock \showarticletitle{GemNet: Menu-Based, strategy-proof multi-bidder auctions through deep learning}.
\newblock \bibinfo{journal}{\emph{arXiv preprint arXiv:2406.07428}} (\bibinfo{year}{2024}).
\newblock


\bibitem[Wang et~al\mbox{.}(2019)]%
        {wang2019learning}
\bibfield{author}{\bibinfo{person}{Weixun Wang}, \bibinfo{person}{Junqi Jin}, \bibinfo{person}{Jianye Hao}, \bibinfo{person}{Chunjie Chen}, \bibinfo{person}{Chuan Yu}, \bibinfo{person}{Weinan Zhang}, \bibinfo{person}{Jun Wang}, \bibinfo{person}{Xiaotian Hao}, \bibinfo{person}{Yixi Wang}, \bibinfo{person}{Han Li}, {et~al\mbox{.}}} \bibinfo{year}{2019}\natexlab{}.
\newblock \showarticletitle{Learning adaptive display exposure for real-time advertising}. In \bibinfo{booktitle}{\emph{Proceedings of the 28th ACM International Conference on Information and Knowledge Management}}. \bibinfo{pages}{2595--2603}.
\newblock


\bibitem[Xie et~al\mbox{.}(2021)]%
        {xie2021hierarchical}
\bibfield{author}{\bibinfo{person}{Ruobing Xie}, \bibinfo{person}{Shaoliang Zhang}, \bibinfo{person}{Rui Wang}, \bibinfo{person}{Feng Xia}, {and} \bibinfo{person}{Leyu Lin}.} \bibinfo{year}{2021}\natexlab{}.
\newblock \showarticletitle{Hierarchical reinforcement learning for integrated recommendation}. In \bibinfo{booktitle}{\emph{Proceedings of the AAAI conference on artificial intelligence}}, Vol.~\bibinfo{volume}{35}. \bibinfo{pages}{4521--4528}.
\newblock


\bibitem[Yan et~al\mbox{.}(2020)]%
        {yan2020ads}
\bibfield{author}{\bibinfo{person}{Jinyun Yan}, \bibinfo{person}{Zhiyuan Xu}, \bibinfo{person}{Birjodh Tiwana}, {and} \bibinfo{person}{Shaunak Chatterjee}.} \bibinfo{year}{2020}\natexlab{}.
\newblock \showarticletitle{Ads allocation in feed via constrained optimization}. In \bibinfo{booktitle}{\emph{Proceedings of the 26th ACM SIGKDD International Conference on Knowledge Discovery \& Data Mining}}. \bibinfo{pages}{3386--3394}.
\newblock


\bibitem[Zhang et~al\mbox{.}(2018)]%
        {zhang2018whole}
\bibfield{author}{\bibinfo{person}{Weiru Zhang}, \bibinfo{person}{Chao Wei}, \bibinfo{person}{Xiaonan Meng}, \bibinfo{person}{Yi Hu}, {and} \bibinfo{person}{Hao Wang}.} \bibinfo{year}{2018}\natexlab{}.
\newblock \showarticletitle{The whole-page optimization via dynamic ad allocation}. In \bibinfo{booktitle}{\emph{Companion Proceedings of the The Web Conference 2018}}. \bibinfo{pages}{1407--1411}.
\newblock


\bibitem[Zhang et~al\mbox{.}(2024)]%
        {zhang2024joint}
\bibfield{author}{\bibinfo{person}{Zhen Zhang}, \bibinfo{person}{Weian Li}, \bibinfo{person}{Yahui Lei}, \bibinfo{person}{Bingzhe Wang}, \bibinfo{person}{Zhicheng Zhang}, \bibinfo{person}{Qi Qi}, \bibinfo{person}{Qiang Liu}, {and} \bibinfo{person}{Xingxing Wang}.} \bibinfo{year}{2024}\natexlab{}.
\newblock \showarticletitle{Joint Auction in the Online Advertising Market}. In \bibinfo{booktitle}{\emph{Proceedings of the 30th ACM SIGKDD Conference on Knowledge Discovery and Data Mining}}. \bibinfo{pages}{4362--4373}.
\newblock


\end{thebibliography}










\end{document}